\documentclass[a4paper,ngerman,UKenglish]{scrartcl}

\usepackage[utf8]{inputenc}
\usepackage{braket,mathrsfs,mathtools,slashed}
\let\originalleft\left
\let\originalright\right
\renewcommand{\left}{\mathopen{}\mathclose\bgroup\originalleft}
\renewcommand{\right}{\aftergroup\egroup\originalright}

\usepackage[normalem]{ulem}

\usepackage{amsmath}
\usepackage{amsfonts}
\usepackage[locale = US]{siunitx} 
\DeclareSIUnit\annum{a}
\usepackage{dsfont}
\usepackage{amssymb,fixmath,upgreek}
\usepackage{textgreek}
\usepackage{lscape,rotating,placeins,graphicx,graphics,multirow,tabularx} 
\usepackage{caption}
\usepackage{subcaption}
\usepackage{nccmath} 
\usepackage{footmisc} 
\usepackage{hyperref}  
\usepackage{enumitem} 
\usepackage{tikz} 
\usepackage{tikzsymbols}
\usepackage{appendix}
\usepackage{lmodern}
\usepackage[newzealand]{babel}
\usepackage{microtype}
\usepackage[T1]{fontenc} 
\usepackage{csquotes} 
\usepackage[sorting=none,doi=true,eprint=true,arxiv=abs,hyperref=true,style=nature-modified,citestyle=numeric-comp]{biblatex}
\DeclareSourcemap{
	\maps[datatype=bibtex,overwrite=true]{
		\map{
			\step[fieldsource=Collaboration, final=true]
			\step[fieldset=usera, origfieldval, final=true]
		}
	}
}

\renewbibmacro*{author}{%
	\iffieldundef{usera}{%
		\printnames{author}%
	}{%
		\printfield{usera}, \printnames{author}%
	}%
}%

\definecolor{lime}{HTML}{A6CE39} 
\newcommand{\orcidicon}{%
	\begin{tikzpicture}
		\draw[lime, fill=lime] (0,0) 
		circle [radius=0.16] 
		node[white] {{\fontfamily{qag}\selectfont \tiny ID}};
		\draw[white, fill=white] (-0.0625,0.095) 
		circle [radius=0.007];
	\end{tikzpicture}
	\hspace{-3mm}
}
\newcommand\orcidMatt{{\href{https://orcid.org/0000-0003-1088-6485}{\orcidicon}}}
\newcommand\orcidSeSc{{\href{https://orcid.org/0000-0003-1997-0026}{\orcidicon}}}
\newcommand\orcidJess{{\href{https://orcid.org/0000-0002-2669-2899}{\orcidicon}}}
\newcommand\orcidJCFe{{\href{https://orcid.org/0000-0003-2441-5801}{\orcidicon}}}

\makeatletter
\newcounter{and}
\newdimen{\instindent}
\newcommand{\institute}[1]{\newcommand{\@institute}{#1}}
\newcommand{\inst}[1]{\unskip\smash{$^{#1}$}\setcounter{and}{1}\ignorespaces}
\newcommand{\email}[1]{\href{mailto:#1}{#1}}
\renewcommand{\maketitle}{
	{
		\raggedright%
		\LARGE%
		\noindent%
		\bfseries%
		\sffamily%
		\@title%
		\newline
		\Large
		\@subtitle
		\par
	}
	
	\vspace{1.5\baselineskip}
	
	{%
		\raggedright%
		\renewcommand{\and}{\unskip, \ignorespaces}%
		\noindent\ignorespaces\@author\par
	}
	
	\vspace{0.5\baselineskip}
	
	{%
		\small%
		\parindent=0pt%
		\parskip=0pt%
		\setcounter{and}{1}%
		\renewcommand{\and}{%
			\par\stepcounter{and}%
			\hangindent\instindent%
			\noindent%
			\hbox to \instindent{\hss\smash{$^{\theand}$\enspace}}\ignorespaces%
		}%
		\setbox0=\vbox{\@institute}%
		\ifnum\value{and}>9\relax\setbox0=\hbox{$^{88}$\enspace}%
		\else\setbox0=\hbox{$^{8}$\enspace}\fi%
		\instindent=\wd0\relax%
		\ifnum\value{and}=1\relax%
		\else%
		\setcounter{and}{1}%
		\hangindent\instindent%
		\noindent%
		\hbox to \instindent{\hss\smash{$^{\theand}$}\enspace}\ignorespaces%
		\fi%
		\ignorespaces%
		\@institute\par
	}
}
\makeatother

\setcapindent{0em} 
\allowdisplaybreaks

\newcommand{\rbr}[1]{\left( #1 \right)} 
\newcommand{\cbr}[1]{\left\{ #1 \right\}}
\newcommand{\sbr}[1]{\left[ #1 \right]} 
\newcommand{\oo}[1]{\frac{1}{#1}}

\newcommand{\defi}{\mathrel{\mathop:}=}

\newcommand{\dif}{\ensuremath{\operatorname{d}}\!}

\newcommand{\LambdaCDM}{{\textLambda\kern-0.06667em CDM} }

\title{Immortality through the dark forces: Dark-charge primordial black holes as dark matter candidates}
\author{Jessica~Santiago\orcidJess\inst{1} \and Justin Feng\orcidJCFe\inst{2} \and Sebastian~Schuster\orcidSeSc\inst{3} \and Matt~Visser\orcidMatt\inst{4}}
\institute{
    Leung Center for Cosmology \& Particle Astrophysics,
	National Taiwan University, Taipei 10617, Taiwan\\E-mail:~\email{{jessicasantiago@ntu.edu.tw}} 
    \and
    Central European Institute for Cosmology and Fundamental Physics, Institute of Physics of the Czech Academy of Sciences,
    Na Slovance 1999/2, 182 21 Prague 8, Czech Republic\\E-mail:~\email{{feng@fzu.cz}}
    \and
	Fysikum, AlbaNova universitetscentrum, Stockholms universitet, 106~91 Stockholm, Sweden\\E-mail:~\email{sebastian.schuster@fysik.su.se}
    \and
	School of Mathematics and Statistics, Victoria University of Wellington, PO~Box~600, Wellington~6140, New~Zealand\\ E-mail:~\email{matt.visser@sms.vuw.ac.nz}}
\date{\today}

\begin{document}
\maketitle
\begin{abstract}

    The fact that no Hawking radiation from the final stages of evaporating primordial black holes {(PBHs)}  has {yet} been observed places stringent bounds on their allowed contribution to dark matter. Concretely, for Schwarzschild PBHs, \emph{i.e.}, uncharged and non-rotating black holes, this rules out black hole masses of less than $10^{-15} M_{\odot}$. 
    In this article, we propose that by including an additional \enquote{dark} $U(1)$ charge one can significantly lower the PBHs' Hawking temperature, slowing down their evaporation process and significantly extending their lifetimes.
    With this, PBHs again become a viable option for dark matter candidates over a large mass range. 
    We will explore in detail the effects of varying the dark electron (lightest dark charged fermion) mass and charge on the evaporation dynamics. For instance, we will show that by allowing the dark electron to have a sufficiently high mass and/or low charge, our approach suppresses both Hawking radiation and the Schwinger effect, effectively extending even the lifespan of PBHs with masses smaller than $10^{-15} M_{\odot}$ beyond the current age of the universe.  
    We will finally present a new lower bound on the allowed mass range for dark-charged PBHs as a function of the dark electron charge and mass, {showing that the PBHs' mass can get to at least as low as $10^{-24}M_\odot$ depending on the dark electron properties}. This demonstrates that the phenomenology of the evaporation of PBHs is ill-served by a focus solely on Schwarzschild black holes.
	\end{abstract}

    \clearpage

    \tableofcontents

    \clearpage

	\section{Introduction}  
    \label{S:Introduction}
	One of the greatest puzzles permeating modern cosmology and the $\Lambda$CDM model regards the nature of the \emph{dark sector}, encompassing both dark matter as well as dark energy. Regarding dark matter candidates, they are usually divided into two broad categories: WIMPs (\textbf{W}eakly \textbf{I}nteracting \textbf{M}assive \textbf{P}articles) and MACHOs (\textbf{MA}ssive \textbf{C}ompact \textbf{H}alo \textbf{O}bjects), such as black holes, rogue planets and ultracompact horizonless objects.
	Each of these categories  has been extensively studied and one can find different constraints imposed by experimental and observational bounds in the available literature. (See references \cite{IceCube:2021xzo, Dhakal:2022rwn, SinghSidhu:2019loh, Egorov:2022fkc} for WIMPs and references \cite{Salvio:2019llz, Graham:2023unf,Brandt:2016aco,Wilkinson:2001vv,Nemiroff:2001bp,Oguri:2017ock,EROS-2:2006ryy,Niikura:2019kqi,LIGOScientific:2019kan} for MACHOs.)
	As a result, these constraints combined impose severe limitations on the possibility of having either MACHOs or WIMPs as the \emph{sole} source of dark matter. 
	In the search for alternatives, many different models have been proposed. 
	These range from more specifically characterizing the dark matter intrinsic properties (\emph{e.g.} \enquote{cold}, \enquote{lukewarm}, \enquote{warm}, \enquote{hot}) to more specific and concrete proposals, such as: axions \cite{Adams:2022pbo}, axion stars \cite{Chang:2024fol}, Q-balls~\cite{Coleman:1985ki, Brihaye:2015veu}, \enquote{strongly} interacting massive particles~\cite{Starkman:1990nj,Mohapatra:1999gg,Hochberg:2014dra}, non-topological solitons~\cite{Brihaye:2015veu, Lee:1991ax, Nugaev:2019vru} and mirror stars~\cite{Sandin:2008db,Curtin:2019ngc, Curtin:2019lhm,Hippert:2021fch}.
	
	Primordial black holes (PBHs), \emph{i.e.}, black holes formed at the end of the inflationary epoch and originated from density fluctuations in the early universe \cite{Ozsoy:2023ryl}, are an ever more commonly suggested type of MACHO. 
    While not being of stellar origin, PBHs are still simply black holes and, therefore, are parametrized (to the best of our knowledge) by the same quantities as their stellar counterparts: Mass $M$, charge $Q$ and angular momentum $J$. A black hole of non-zero mass and of non-zero charge (but of zero angular momentum) is described by the Reissner--Nordström (RN) metric. 
    Note that, even though the charge is normally assumed to be of (standard) electromagnetic origin, the only requirement to regain the RN metric as a valid description of such black holes is for the Einstein equation to be coupled to \emph{some} (classical) $U(1)$ gauge theory described by Maxwell's equations. This may include an additional \enquote{magnetic} charge, but we will ignore this possibility in our present paper. 
 
\begin{figure}[!ht]
		\centering
		\includegraphics[scale=0.75]{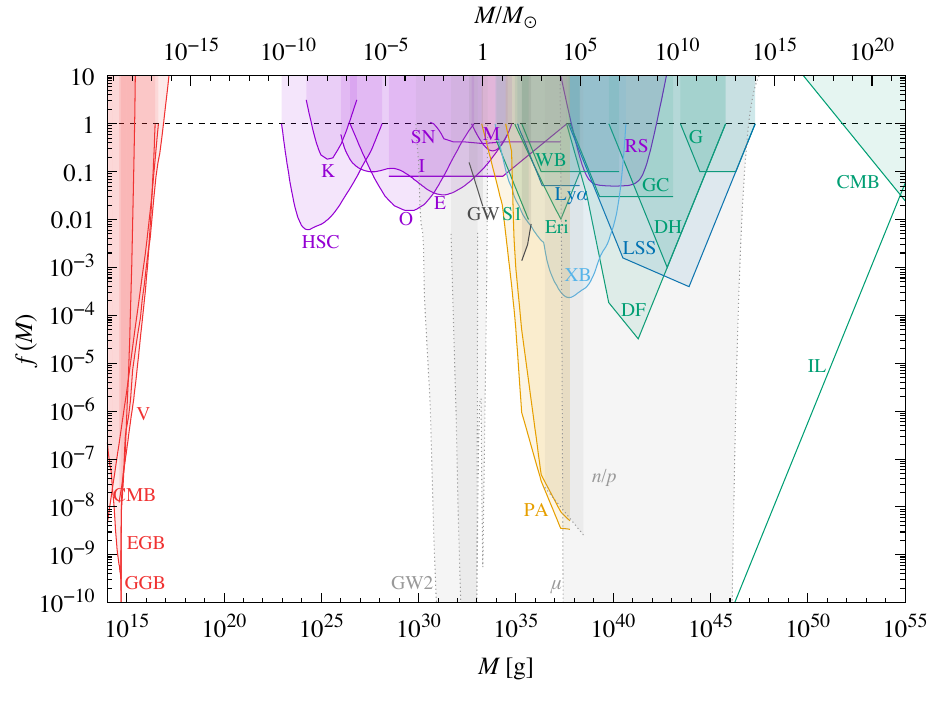}
		\caption{\small{Reprinted figure with permission from \cite{Carr:2020gox}. Original caption: \enquote{Constraints on $f(M)$ from evaporation (red), lensing (magenta), dynamical effects (green), gravitational waves (black), accretion (light blue), CMB distortions (orange), large-scale structure (dark blue) and background effects (grey). Evaporation limits come from the extragalactic $\gamma$-ray background (EGB), CMB anisotropies (CMB), the Galactic $\gamma$-ray background (GGB) and Voyager-1 $e^{\pm}$ limits (V). Lensing effects come from microlensing of stars in M31 by Subaru (HSC), in the Magellanic Clouds by MACHO (M) and EROS (E), in the local neighbourhood by Kepler (K), in the Galactic bulge by OGLE (O) and the Icarus event in a cluster of galaxies (I), microlensing of supernovae (SN) and quasars (Q), and millilensing of compact radio sources (RS). Dynamical limits come from disruption of wide binaries (WB) and globular clusters (GC), heating of stars in the Galactic disk (DH), survival of star clusters in Eridanus II (Eri) and Segue 1 (S1), infalling of halo objects due to dynamical friction (DF), tidal disruption of galaxies (G), and the CMB dipole (CMB). Accretion limits come from X-ray binaries (XB), CMB anisotropies measured by Planck (PA) and gravitational waves from binary coalescences (GW). Large-scale structure constraints come from the Lyman-A forest (Ly$\alpha$) and various other cosmic structures (LSS). Background constraints come from CMB spectral distortion (µ), 2nd order gravitational waves (GW2) and the neutron-to-proton ratio ($n/p$). The incredulity limit (IL) corresponds to one hole per Hubble volume.}}}
		\label{fig:PBH4}
	\end{figure}

	Constraints on the fraction of dark matter composed of \emph{uncharged} PBHs $(f_{\textrm{PBH}})$, come from observations of gravitational waves \cite{Kavanagh:2018ggo}, the cosmic microwave background (CMB) \cite{Serpico:2020ehh}, microlensing \cite{Niikura:2019kqi, Smyth:2019whb}, the {21}{cm} absorption line \cite{Hektor:2018qqw}, gas heating \cite{Lu:2020bmd}, radio and X-ray emissions \cite{Manshanden:2018tze}, ultra-faint dwarf galaxies \cite{Stegmann:2019wyz,}, and more. These constraints vary significantly with the primordial black hole mass $M$. In figure~\ref{fig:PBH4}, reproduced from \cite{Carr:2020gox} with the authors' permission, one can find recent observational constraints on $f_{\textrm{PBH}}$ for Schwarzschild black holes. 
	
	Due to their weak gravitational effects, low mass PBHs are naturally harder to detect and, as can be seen in figure~\ref{fig:PBH4}, for PBHs with masses $M_{\textrm{PBH}}$ below $\sim 10^{-15}M_{\odot}$, the constraints come solely from \emph{the lack of (direct and indirect) observation} of their Hawking evaporation \cite{Carr:2020gox,Luo:2020dlg, Keith:2020jww, Korwar:2023kpy, Boudaud:2018hqb, Auffinger:2022khh, HESS:2023zzd}.
    Were such sufficiently low mass PBHs formed in the early universe, their Hawking temperature would have been so high that they should have evaporated by now. For distant PBHs we therefore should be able to still observe these evaporation events. Furthermore, the particles expected to be emitted at the final stages of the evaporation process should have left signals in the observed early universe data \cite{Carr:2020gox}.
    Given the absence of such signals, we are left with three possible conclusions: (1) Low mass PBHs were formed, but we are still unable to observe their evaporation and to correctly infer the signals such evaporation would have left in the early universe; (2) Such PBHs were formed, but their evaporation process was halted for some reason; (3) Such PBHs simply did not form in the first place. In this paper, we will focus on the effects of option (2).
  
    On the other hand, low mass PBHs are the only dark matter candidate which can still account for a significant fraction or \emph{even the totality} of the dark matter in the universe without resorting to stable new particles still present to the current epoch \cite{Montero-Camacho:2019jte,DeLuca:2020agl}. Hence, exploring the idea of PBH remnants, and mechanisms which can stop --- or at least significantly slow down --- their evaporation process, is of great importance in order to put updated constraints on PBH as dark matter candidates. This is the aim of this work. 

    For this, we shall consider \emph{charged} PBHs. It is a well known fact that, when a black hole approaches extremality, in the RN case meaning when the charge-to-mass ratio $Q/M$ approaches unity, the Hawking temperature $T$ tends to zero:
	\begin{equation}
		M \to Q \;\Longrightarrow\; T \to 0.
	\end{equation}
	If a black hole has a standard electromagnetic charge, this will not be a solution: Astrophysical, electromagnetically charged (and non-spinning) black holes have the \enquote{bad habit} of neutralizing extremely fast through charge accretion from surrounding material 
	\cite{Eardley:1975kp,Gong:2019aqa}. This alone already imposes extremely strong limitations on the possible charge-to-mass ratio an electromagnetic RN black hole may have. In terms of the electron charge $e$ and electron mass $m_e$, this is:
	\begin{equation}
		\frac{Q}{M} \ll \frac{m}{e} \simeq 10^{-21}\qquad \textrm{(in geometrodynamic units)}\;.
	\end{equation}
    Therefore, in any attempt to slow down the Hawking evaporation process by approaching extremality, standard electromagnetic charge must be clearly ruled out. Yet even from this point of view, already small, residual charges would have significant observational impact \cite{Zajacek:2019kla}.
    
    Moreover, given that we are interested in looking at black holes in the mass range $\lesssim 10^{-15} M_{\odot}$, the problem does not arise only from charge accretion, but also from the fact that the smallest electromagnetically charged particle in the standard model of particle physics, the electron $e^-$, is much too light. As will be shown in the following sections, this factor, combined with the fact that the electron charge itself is relatively high, causes low mass black holes to also rapidly discharge via both Hawking effect and Schwinger pair production.
	
    The solution we propose here is to give the PBH a \enquote{dark electromagnetic} $U(1)$ charge. We will further discuss the dark electromagnetism model adopted by us in section~\ref{S:DarkE}. Essentially, by assuming the lightest dark charged particle (the \enquote{dark electron}) to be heavier than the standard model electron, one is able to freeze the Hawking evaporation process for specific mass ranges of PBHs --- even when the initial black hole charge-to-mass ratio $Q/M$ is much smaller than one. In sections~\ref{s:parameter space} and~\ref{S:Muniv} we will present how the dark electron charge $e_\chi$ and mass $m_\chi$ play a crucial role in determining the PBH mass ranges in which near-extremality can be achieved (via evaporation). This will allow us to introduce a new lower limit mass for darkly charged PBHs, above which the expected lifetime of the PBHs is still larger than the age of the universe --- extending the Hawking radiation constraints, currently valid only for Schwarzschild.
    
    The idea of using near-extremality of black holes to increase their potential as dark matter candidates was employed before: In the context of the weak gravity conjecture encountered in string theory, a very closely related study is that of reference \cite{Bai:2019zcd}. We will not assume the validity of the weak gravity conjecture in our model, though our numerical values will still conform to it. In the context of rotating, uncharged black holes, reference \cite{Taylor:2024fvf} suggested a specific toy model. Likewise, (P)BHs have been used before to constrain dark matter models themselves \cite{Dai:2009hx,Abbasvandi:2016oyw,Cardoso:2018tly,Kitabayashi:2022uvu,Gupta:2023voq,Gu:2023eaa,Kim:2024iud}. On a different note, quantum effects such as \enquote{memory burden}~\cite{Dvali:2018xpy,Dvali:2020wft,Dvali:2024hsb,Thoss:2024hsr,Alexandre:2024nuo,Barker:2024mpz}, as well as those arising from the postulated discreteness of states near extremality \cite{Preskill:1991tb,Maldacena:1998uz,Page:2000dk,Brown:2024ajk}, may also have an impact on the evaporation rate of PBHs. The memory burden was also considered in the context of charged black holes and dark matter particles before \cite{Barman:2024kfj}, albeit without correctly accounting for the Schwinger effect as done in the present article. With regard to the latter, we find that our solutions remain far from the threshold required to significantly modify the evaporation timescales (see Appendix~\ref{S:numerics}).     
    Moreover, being close to extremality allows for several significant simplifications in modeling the Hawking evaporation of RN black holes. Concretely, links to calculations in anti-deSitter space-times that have already achieved textbook status \cite{Fabbri:2005mw} can be made.

    The paper is organized as follows: In section~\ref{S:Evap-RNBH} we introduce the RN metric and discuss the main results and limitations of earlier work by Hiscock and Weems (HW) \cite{Hiscock:1990ex} on charged evaporating black holes. In section~\ref{S:Beyond}, we analyze, modify and extend the results on charged evaporating black holes. We also include estimations for the approximate evaporation times in three different regimes: near-extremal, small charge limit and along the so-called attractor curve (or attractor region). In section~\ref{S:DarkE} we update the original HW limitations to the case of dark electromagnetism, highlighting the role played by varying the dark electron properties. In section~\ref{S:near_extr_time} we update the evaporation time estimates to dark-charged black holes. In section~\ref{S:Muniv} we present the updated lower bounds for the mass of dark-charged PBHs to live longer than the age of the universe. We explicitly analyze different evaporation regimes and present the final results for different dark electron mass $m_\chi$ and charge $e_\chi$.    
    In Appendix~\ref{S:numerics} we discuss the numerical methods used, together with possible issues which come from the stiffness of the differential equations encountered.  In section~\ref{S:conclusions} we present the final discussions and conclusions. The numerical codes and scripts we employed for the HW model can be found on the Github repository \url{https://github.com/justincfeng/bhevapsolver}.
    
    \paragraph{Conventions and notation:} Geometrodynamic units are used throughout, unless explicitly stated otherwise. See appendix~\ref{app:units} for details. The metric signature is \enquote{mostly positive}, $(-+++)$. 
	
	\section{Evaporating Reissner–Nordström Black Holes}\label{S:Evap-RNBH}
    One of the most important and prominent work dealing with the evaporation process of charged RN black holes was presented by Hiscock and Weems in 1990~\cite{Hiscock:1990ex}. In this section, we will briefly present their main results, the limitations of their work and how we will extend it beyond such limitations.

    \subsection{Reissner--Nordström metric}
	The Reissner--Nordström metric is given in geometrodynamic units by:
	\begin{equation}\label{eq:RN}
		\dif s^2 = -\left(1-\frac{2M}{r}+\frac{Q^2}{r^2}\right)\dif t^2 + \left(1-\frac{2M}{r}+\frac{Q^2}{r^2}\right)^{-1} \dif r^2 + r^2 \dif \Omega^2, 
	\end{equation}
	where (as before) $M$ and $Q$ are the mass and charge of the black hole, respectively, and $\dif \Omega^2$ is the line element of the 2-sphere. This metric:
	\begin{itemize}
		\item Is static, spherically symmetric;
		\item Solves the coupled Einstein--Maxwell equations;
		\item Has two event-horizons, outer and inner, located at:
		\begin{equation}
			r_\pm = M \pm \sqrt{M^2 - Q^2}.
		\end{equation}
	\end{itemize}
    Note that throughout this article we will restrict ourselves to the case $Q<M$. Although we will discuss near-extremal situations, extremality will not be required, achieved or desired in any step of our calculations.\footnote{Therefore, we can and will completely avoid discussions regarding the weak-gravity conjecture, which is very much outside the scope of this article \cite{Harlow:2022ich}. Nonetheless, it has informed very closely related studies on extremal PBHs as dark matter candidates \cite{Bai:2019zcd}.}

	The Hawking temperature of an evaporating RN black hole is given by:
	\begin{equation}
		T = \frac{\hbar \kappa}{2\pi}, \qquad \textrm{with} \qquad \kappa = \frac{(M^2 - Q^2)^{1/2}}{r_+^2}\;.
        \label{eq:Temp}
	\end{equation}
	Here, $\kappa$ is the surface gravity (of the outer horizon) of the black hole. This is the temperature encountered in the semiclassical derivation of the Hawking effect.

	\subsection{Evolution equations}
	The evaporation process of charged black holes, as discussed in Hiscock's and Weems' beautiful, original paper \cite{Hiscock:1990ex}, is driven by two distinct quantum effects: the Hawking effect and the Schwinger effect. The Hawking effect can be linked to the non-uniqueness of a vacuum state in a general space-time (and in particular to the difference of a vacuum close to a horizon and a vacuum at asymptotic infinity). The Schwinger effect is due to the separation of particle-antiparticle pairs of vacuum fluctuation in a strong electromagnetic background field. While sometimes the Hawking effect is likened to a gravitational analogue of the Schwinger effect, the role of the horizon (or a surface close to one) limits the usefulness of this analogy.
	
	In the charged black hole context, HW have argued that while a RN black hole's mass loss occurs through the emission of both charged and uncharged particles --- \emph{i.e.}, via both effects --- the loss of charge is dominated by the Schwinger effect. The resulting coupled ordinary differential equations describing the time evolution of the mass $M$ and charge $Q$ of such black holes are given by:
	\begin{align}
		\frac{\dif M}{\dif t} &= -a T^4 \alpha \sigma_0 + \frac{Q}{r_+} \frac{\dif Q}{\dif t}, \label{eq:dMdtHWfull}\\
		\frac{\dif Q}{\dif t} &= -\frac{e^3}{\pi^2 \hbar^2 r_+} \exp\rbr{-\frac{r_+^2}{QQ_0}} - \frac{\pi}{\sqrt{QQ_0}} \operatorname{erfc}\rbr{\frac{r_+}{\sqrt{QQ_0}}}, \label{eq:dQdtHWfull}
	\end{align}
	where $a=\pi^2/(15\hbar^3)$, 
    and $\alpha$ is a numerical correction factor (not the fine structure constant) for the scattering cross-section of all massless particles involved (for three neutrinos $\alpha \approx \num{2.0228}$ and $\alpha \approx \num{0.2679}$ for no neutrinos).
     Furthermore $\sigma_0$ is the idealized, geometric optics scattering cross-section, given by\footnote{In terms of the outer (unstable) circular photon orbit $r_{\gamma+} \defi \frac{3M}{2}
    \rbr{1+\sqrt{1-\frac{8}{9}Q^2}}$ \cite{Slezakova:2006PhD,Cederbaum:2015fra}, note that equation~\eqref{eq:sigma0} can be re-expressed as
		\begin{equation*}
			\sigma_0 = \pi \frac{r_{\gamma+}}{8\rbr{r_{\gamma+}M-4Q^2}}.
		\end{equation*}
		While more compact, in the following we will maintain the full expression with the explicit dependence on $M$ and $Q$.}:
	\begin{equation}
		\label{eq:sigma0}
		\sigma_0 \defi \pi\frac{\rbr{3M + \sqrt{9M^2 -8Q^2}}^4}{8\rbr{3M^2-2Q^2+M\sqrt{9M^2-8Q^2}}}\;,
	\end{equation}
     Note that in all the analysis that follows, we will assume the number of neutrinos to be zero. This choice is justified by the fact that, as previously shown in \cite{Hiscock:1990ex}, while the number of neutrinos has an impact on the precise time-scale of the system's evolution, we have verified that the order of magnitude of the evaporation time is the same for both cases of zero or three neutrinos. Moreover, the other considered aspects of the evolution are not affected by the neutrino number. Straightforward checks confirmed that this holds true for charged and dark-charged black holes. 

	Furthermore, 
	\begin{equation}
		Q_0 \defi \frac{\hbar e}{\pi m^2} \approx \num{1.7e5} M_\odot \label{eq:q0}
	\end{equation}
	is a \enquote{charge mass scale} naturally encountered in the context of the Schwinger effect \cite{Schwinger:1951nm,Zaumen:1974nat,Gibbons:1975kk,Hiscock:1990ex}. While this is fixed in the standard model of particle physics, it will change for alternative $U(1)$ theories. We remind the reader that geometrodynamic units have been adopted throughout.

	\subsection{The Hiscock--Weems (HW) model: assumptions and limitations} \label{S:assumptions} 
	
    In this subsection, we will discuss the various assumptions and approximations needed to arrive at the evolution equations~\eqref{eq:dMdtHWfull} and~\eqref{eq:dQdtHWfull}.\footnote{While \cite{Gibbons:1975kk} heavily informed the analysis of \cite{Hiscock:1990ex}, its assumptions and conjectures are not very clearly listed, documented or delineated. Concretely, the analysis of greybody factors has only recently become analytically tractable \cite{Bonelli:2022ten,Lisovyy:2021bkm}, and for applications of these results to charged black holes, see \cite{Cavalcante:2021scq,Amado:2021erf}. For clarity's sake, we will therefore follow primarily \cite{Hiscock:1990ex}.} Our overall goal is to use the analysis of HW~\cite{Hiscock:1990ex} as starting point for considering near-extremal, dark-charge PBHs as dark matter candidates. The possible mass range of such dark matter PBHs depends on the parameters $e_\chi$ and $m_\chi$ of our dark $U(1)$ model, so it is important to check how these parameters are constrained by the assumptions behind the HW model. This will be undertaken later, in section~\ref{S:updated-cond}. 
    For now, we will list the assumptions made, the physical reasons supporting them, then we quickly state the resulting inequality, before summarizing the argument behind it.
	\begin{itemize}
		\item \textbf{Positive charge:}
			\begin{equation}
				Q>0.\label{eq:posQ}
			\end{equation}
			This is simply a convention that greatly simplifies the notation. Due to the charge-symmetry of the problem, all main results stay the same if one decides to assume negative charges.
		\item \textbf{Applicability of Schwinger’s
result:}
			\begin{equation}
				M\gg \frac{\hbar}{m_e} \simeq 10^{-15}M_{\odot}\;.\label{eq:constr1}
			\end{equation}
			The black hole mass is always assumed much larger than the reduced Compton wavelength of the electron (or lightest charged particle). This imposes a lower bound on the mass of the black holes analysed. Under this restriction, one may then use the flat-space expression of Schwinger \cite{Schwinger:1951nm} for the rate of electron-positron pair creation per unit four-volume~\footnote{Note that the {use of} four-volume implies {the use of} proper time $\tau$, while everything in the following analysis is in terms of coordinate time $t$.}:
			\begin{equation}
				\Gamma = \frac{(eE)^2}{4\pi^3\hbar^2 c}\sum_{n=1}^{\infty} \frac{1}{n^2}\exp{\rbr{-\frac{\pi m_e^2 c^3 n}{\hbar eE}}}\;.
				\label{eq:Schwinger}
			\end{equation}
            {Note that we have recovered SI units in this particular equation.}
			\item \textbf{Series truncation:}
			\begin{equation}
				\frac{e^3 Q}{m_e^2 r^2} \ll 1
				\label{eq:approx1}
			\end{equation}
			This follows HW, in order to truncate the Schwinger pair creation rate~\eqref{eq:Schwinger} after the first term of the series.
			\item \textbf{Weak field limit:}
			\begin{equation}
				M \gg \frac{e^3}{m_e^2} \approx \num{4e3} M_{\odot}.
				\label{eq:constr2}
			\end{equation}
			As the original Schwinger effect is a flat-space-time calculation, for equation~\eqref{eq:approx1} to be valid in the entire domain of outer communication (that is, $r>r_+$), one must impose what essentially comes down to the above \emph{weak field limit} condition.
			\item \textbf{Error function series truncation:}
			\begin{equation}
				r_+^2 \gg QQ_0\;.
				\label{eq:rplus-gg-QQ0}
			\end{equation}
			This ensures that the error function in equation~\eqref{eq:dQdtHWfull} is well approximated by the first term(s) of its asymptotic series. 
            Note that a simplified, sufficient but not necessary condition is $M\gg Q_0$, meaning that inequality~\eqref{eq:rplus-gg-QQ0} will always be satisfied for large enough masses. In the 
            analysis presented in HW \cite{Hiscock:1990ex}, this implies that the results are certainly valid for any black hole with a mass greater than $\approx 10^6 M_\odot$. In section~\ref{S:updated-cond} we will show how this condition is actually much less restrictive than the simplified bound imposed by HW.
	\end{itemize}
	
	\paragraph{Summarized model limitations:} When combined, all of the above conditions can be summarized as:
	\begin{equation}
		\textrm{(i)}\;\; M \gg \frac{\hbar}{m_e}, \qquad \qquad 
        \textrm{(ii)}\;\; \frac{e^3 Q}{m_e^2 r^2} \ll 1, \qquad \textrm{and} \qquad \textrm{(iii)}\;\;r_+^2 \gg QQ_0
        \;.
        \label{eq:summ_conditions}
	\end{equation}
	We will revisit these conditions in section~\ref{S:updated-cond}, where we will investigate their impact on the parameter space of our chosen \enquote{dark} $U(1)$ model.

    \subsection{Main results of the HW model}\label{S:HWresults}
    In this section, we will briefly discuss the main results regarding the evaporation of charged black holes obtained by Hiscock and Weems in \cite{Hiscock:1990ex}. 

    \paragraph{The $(M,(Q/M)^2)$ configuration space}
    
\begin{figure}[!ht]
    \centering
    \includegraphics[width=0.65\linewidth]{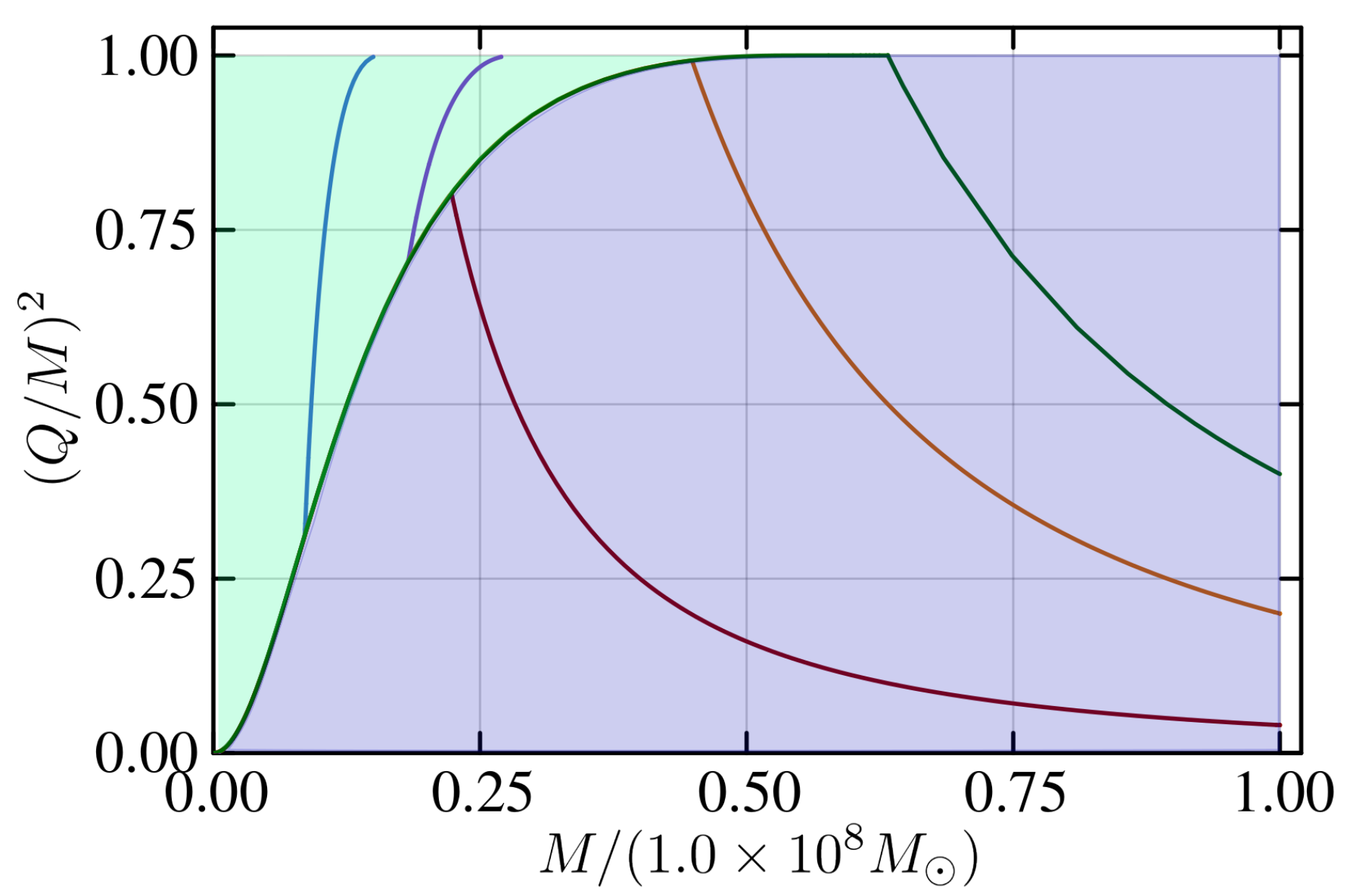}
	\caption{Recreation of figure~2 \emph{of} \cite{Hiscock:1990ex}, showing the evolution of $(Q/M)^2$ and $M$ in configuration space for different, initial black hole masses and charges, $(M,Q)$. The shaded blue region represents the \enquote{mass dissipation zone}, while the shaded light green region represents the \enquote{charge dissipation zone}.} 
	\label{fig:configuration-space}
    \end{figure}

    {HW present a particularly surprising result concerning the configuration space evolution of an evaporating charged black hole. In figure~\ref{fig:configuration-space} we recreate figure 2 from the original HW paper. Curves in this plot represent the evolution of black holes with different initial mass and charge $(M, Q)$ in the configuration space. 
    Given that (in the absence of accretion) the black hole mass is monotonically decreasing, the direction of time flows towards lower mass (to the left) in each curve. However, as will be shown, the amount of time a black hole will persist in the neighbourhood of each point of the configuration space may vary by several orders of magnitude.} 

    \enlargethispage{20pt}
    One may immediately identify two distinct regions in the phase diagram, which (following the terminology of \cite{Hiscock:1990ex}) are termed the \enquote{mass dissipation zone} and \enquote{charge dissipation zone}, presented in figure~\ref{fig:configuration-space} in shaded colors blue and light green, respectively. These two regions are distinguished by the behavior of the squared charge-to-mass ratio $(Q/M)^2$; under time evolution (recalling that mass decreases with time), the mass dissipation zone is characterized by curves with an increasing $(Q/M)^2$, and the charge dissipation zone is characterized by a decreasing $(Q/M)^2$. The mass dissipation zone and charge dissipation zone are separated by a third region, which we shall refer to as an \enquote{attractor region} as an extension of HW's terminology of the \enquote{attractor curve}. The attractor region is a narrow basin of attraction in which the curves from both the charge and mass dissipation zones accumulate.  
    We will further explore the features and properties of the attractor region in section~\ref{S:Beyond}, after we have introduced the nondimensionalized form of equations~\eqref{eq:dMdtHWfull} and~\eqref{eq:dQdtHWfull}.

	In order to understand {the reasons leading to the split of the configuration space in different regions}, one must recall that due to the presence of charge, both the Hawking as well as the Schwinger effect are responsible for driving the evaporation process. 
    {The coexistence of those distinct competing effects then manifests as a sharp split in the configuration space.    
    In this way, while the evaporation of black holes in the {mass dissipation zone} is primarily driven by the Hawking effect, in the \enquote{charge dissipation zone} the exponentially quick charge loss is driven solely by the Schwinger effect}~\footnote{Note that due to the extremely low BH temperatures, in their work, HW assumed that all mass lost in the Hawking process is due to the emission of massless (and therefore uncharged) particles. This implies that no charge-loss ever occurs due to Hawking radiation. The justification lies on the fact that, even though a black hole may still loose some charge due to the emission of charged particles, the particle production rate will be exponentially suppressed by the mass of the lightest charged particle --- in their case, the electron mass.\label{footnote}}. As mentioned, in section~\ref{S:Beyond} all the details of such dynamical processes will be dissected. For now, we would simply like to point out that, in the absence of charge-loss due to accretion and interaction with surrounding matter, a black hole with a sufficiently large initial mass may evolve into near-extremality even when its initial $Q/M$ ratio is very low.

    \enlargethispage{20pt}
    A closer look at the configuration space will also reveal a region of positive specific heat. We include a short discussion of this region in appendix~\ref{app:spec-heat}. While this is an often overlooked feature of black holes beyond the Schwarzschild solution, it plays little role in the evaporation process itself. This becomes clear when one compares the various analyses of the upcoming sections. For example, by comparing figure~\ref{fig:configuration-space} with figure~\ref{fig:HW1}, it is clear that no noticeable change in the trajectories coincides with this region of the configuration space.

     \begin{figure}[ht!]
		\begin{subfigure}{0.48\textwidth}
			\includegraphics[width=\linewidth]{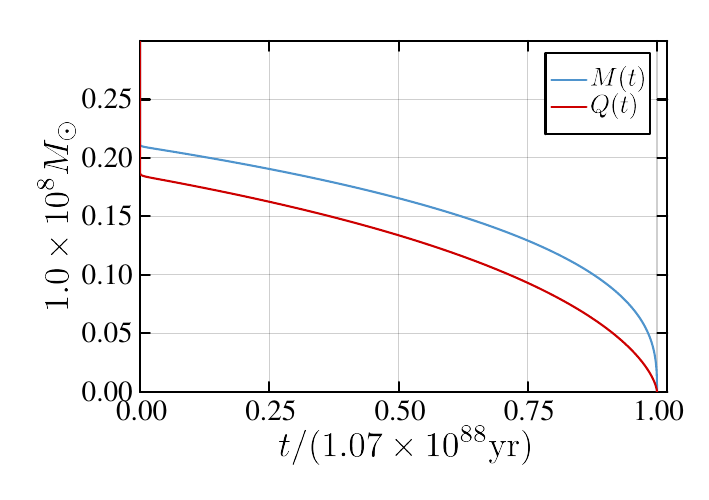}
			\caption{}
			\label{fig:evaporation1a}
		\end{subfigure}\hspace*{\fill}
	\begin{subfigure}{0.48\textwidth}
			\includegraphics[width=\linewidth]{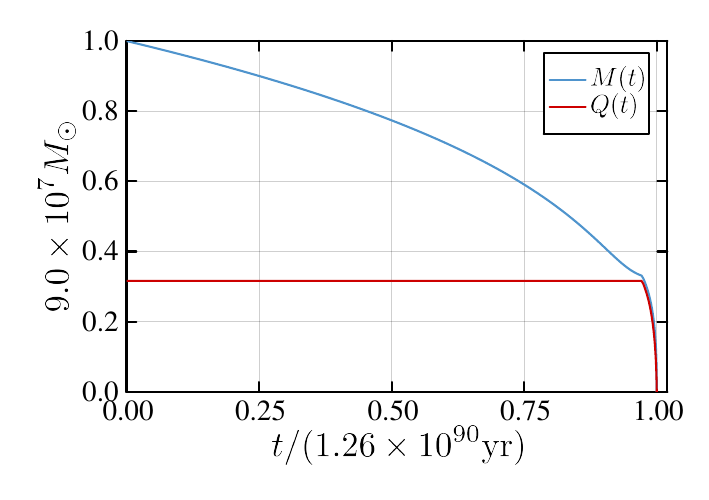}
			\caption{}
			\label{fig:evaporation1b}
		\end{subfigure}
  \caption{On the left, the evolution of $M(t)$ and $Q(t)$ in the charge dissipation zone are illustrated, with an initial mass of $M(0)=\num{0.3e6} M_{\odot}$ and an initial charge $Q(0)=\num{0.99e6} M_{\odot}$, qualitatively recreating the results of figure~3 in \cite{Hiscock:1990ex}. On the right is a recreation of figure~5 of \cite{Hiscock:1990ex}, with $M(0)=\num{9.0e7} M_{\odot}$ and $Q(0)=\sqrt{0.1} M(0)$.}
  \label{fig:evaporation1}
	\end{figure}

\begin{figure}[hb!]
		\begin{subfigure}{0.48\textwidth}
		 	\includegraphics[width=\linewidth]{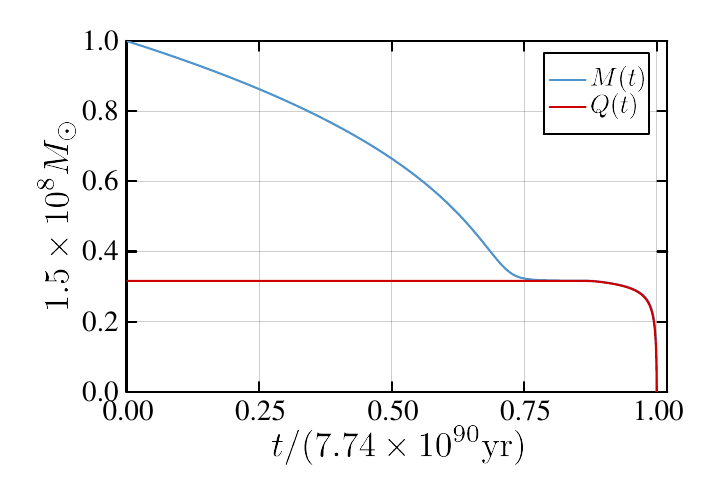}
		 	\caption{}
		 	\label{fig:evaporation2a}
		\end{subfigure}\hspace*{\fill}
		\begin{subfigure}{0.48\textwidth}
			\includegraphics[width=\linewidth]{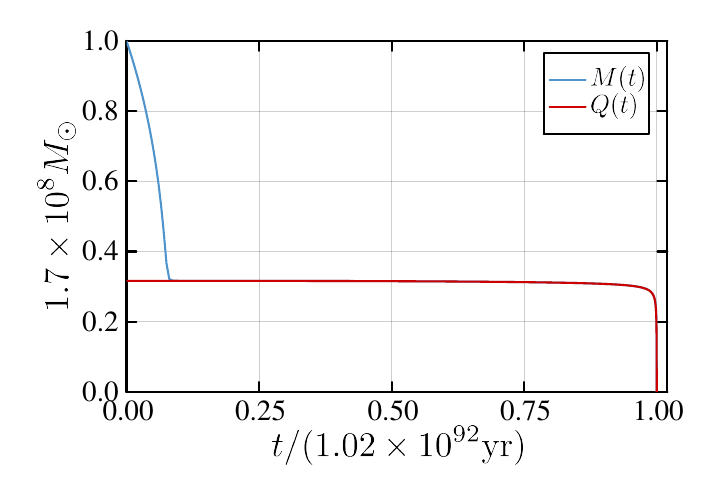}
			\caption{}
			\label{fig:evaporation2b}
		\end{subfigure}
            \caption{On the left is a recreation of figure~6 of \cite{Hiscock:1990ex}, with $M(0)=\num{1.50e8} M_{\odot}$ and $Q(0)=\sqrt{0.1} M(0)$. On the right is a recreation of figure~7 of \cite{Hiscock:1990ex}, with $M(0)=\num{1.68e8} M_{\odot}$ and $Q(0)=\sqrt{0.1} M(0)$.  
            }
            \label{fig:evaporation2}
    \end{figure}

    \paragraph{On the evaporation timescales}
    Using numerical methods, HW also provide the total evaporation time for black holes with masses ranging from $10^6 M_{\odot}$ to $10^8 M_{\odot}$  and a variety of initial $Q/M$ configurations.
    The necessary, underlying assumption is that equations~\eqref{eq:dMdtHWfull} and~\eqref{eq:dQdtHWfull} remain valid throughout the numerical evolution. We repeated HW's numerical analysis as a simple consistency check for our later extensions. For details on the numerics, see Appendix~\ref{S:numerics} below. Our results are shown in figures~\ref{fig:evaporation1} and~\ref{fig:evaporation2}. 

    Comparing the evaporation timescales of figure~\ref{fig:evaporation2b} and~\ref{fig:evaporation1a}, one sees that keeping the initial charge-to-mass ratio,  an increase of the initial mass by a mere factor of 8 already prolongs the evaporation time by roughly $10^5$ years.
    This is due to the presence of a \enquote{plateau} on which the black hole evaporation enters a \enquote{semi-frozen} state. As long as the black hole state is on this plateau, virtually no mass and virtually no charge are lost until the plateau is (after a very long time) exited again. 
    The presence or absence of such a plateau depends on where in the configuration space a particular black hole is initially located --- or more precisely, 
    whether it approaches near-extremality via the evaporation process or not.
    
    The idea is conceptually simple: the temperature of a black hole tends to zero as $Q \to M$ --- meaning the Hawking evaporation process is slowed down by a power of $T^4$. On the other hand,
    the Schwinger effect is exponentially suppressed in the plateau region. A more thorough investigation of the evaporation behaviour and associated timescales will be given in section~\ref{S:evap_time_est}.

    \subsection{Extending the HW analysis}
    
    As can be seen in figure~\ref{fig:configuration-space}, the mass range where the attractor region is located for black holes with a standard electromagnetic charge is of the order of $10^7\;M_\odot$.
    This implies that only black holes with masses greater than $\sim10^6\;M_\odot$ will exhibit the interesting \enquote{life-extending} features that arise from approaching near-extremal states.
    This is limiting \textit{per se}, but it becomes even more crucial when talking about primordial black holes. 
    As previously stated, our goal is to extend the range of masses for PBHs consistent with observations (as given in figure~\ref{fig:PBH4}). Therefore it makes sense to focus on the mass region below $10^{-15}\;M_{\odot}$ --- as this is excluded for Schwarzschild PBHs due to Hawking radiation constraints, whereas RN PBHs may still remain as a valid option. However, to do so, it is clear one must find a way to \enquote{push} the attractor region into the low mass regimes.
     
    As will be clarified on section~\ref{S:DarkE}, the location of the attractor is fixed by the scales of the forces present in the system --- in the standard electromagnetism case, the electron mass $m$ and charge $e$. In light of this, in order to explore the low black hole mass regime, one must demand the black hole charge to be sourced by a different interaction than standard electromagnetism. For simplicity's sake, we shall work with a $U(1)$ dark electromagnetism sector, consisting of massless dark photons and a single, massive, charged particle --- what we will call the \enquote{dark electron}. An in-depth discussion of the model and the results obtained will be presented from section~\ref{S:DarkE} onward.

    In the following section we focus primarily on the case of standard electromagnetism, even though the qualitative results are kept unaltered by changes in the properties of the dark electron. We shall, however, rewrite the evolution equations in a more suitable way which will help us better understand the underlying dynamics --- which will be useful when finally moving into the dark sector.

    \section{Beyond the Hiscock and Weems analysis}\label{S:Beyond}
    
    \subsection{Modified equations}
    \label{S:modified}
    It is convenient to rewrite equations~\eqref{eq:dMdtHWfull} and~\eqref{eq:dQdtHWfull}, for the time evolution of mass and charge, in terms of the dimensionless variables $Y\defi(Q/M)^2$ and $\mu\defi M/M_\textrm{s}$, where $M_\textrm{s}$ is a newly introduced mass scale. It is chosen and serves to keep all of the system's behaviour and nontrivial features of interest to us confined to the range $\mu\in(0,1)$. Equation~\eqref{eq:dMdtHWfull} can now be written in terms of the new variables as:
	\begin{align}
	    \frac{dM}{dt} &= -a T^4 \alpha \sigma_0 + \frac{Q}{r_+} \frac{dQ}{dt}, \\    
	    \Longleftrightarrow\quad \frac{M_{\textrm{s}}d\mu}{dt}&=\frac{\alpha \hbar}{1920 M_\textrm{s}^2 \pi}\frac{\left(\sqrt{9-8 Y}+3 \right)^4 (Y-1)^2 }{\mu^2 \qquad \left(\sqrt{1-Y}+1\right)^8 \left(3 -2 Y+\sqrt{9-8Y}\right)}
	    \\ \nonumber
	   &\hphantom{=} -\frac{e^4}{2 m_e^2 \hbar \pi^3} \frac{Y^2}{\left(\sqrt{1-Y}+1\right)^4} \exp{\left(-\frac{\pi m_e^2 M_\textrm{s}}{e \hbar}\frac{\mu\left(\sqrt{1-Y} +1\right)^2}{\sqrt{Y}}\right)}.
	\end{align}
   The first term comes from the Hawking evaporation contribution and the second from the Schwinger effect. In this way, it is convenient to rewrite $\dot{M}$ in terms of the \enquote{Hawking-associated} $H(\mu,Y)$ and \enquote{Schwinger-associated} $S(\mu,Y)$ functions: 
    \begin{equation}
	H(\mu,Y)\defi\frac{\left(\sqrt{9-8 Y}+3\right)^4 (1-Y)^2}{\mu ^2 (\sqrt{1-Y}+1)^4 (3-2 Y+\sqrt{9-8 Y})},\label{eq:Haw}
	\end{equation}
    \begin{equation}
		S(\mu,Y)\defi
        \exp\left\{b_0\left[z_0-{\mu \left(\sqrt{1-Y}+1\right)^2}/{\sqrt{Y}}\right]\right\}.
        \label{eq:S-eq}
	\end{equation}
    Here, the dimensionless constants $s_0$, $z_0$, and $b_0$ are given by:
    \begin{equation}
    \label{eq:ConstDefs}
         s_0=\frac{\alpha  \hbar }{1920 \pi  M_\textrm{s}^2}
        ,\qquad 
        z_0=\frac{e \hbar}{\pi m_e^2 M_\textrm{s}}\ln\left(\frac{960 e^4 M_\textrm{s}^2}{\pi ^2 \alpha m_e^2 \hbar ^2}\right)
        ,\qquad
        b_0=\frac{m_e^2 M_\textrm{s} \pi}{e \hbar}.
	\end{equation}
    Note that, in terms of the \enquote{charge mass scale} $Q_0$ introduced in equation~\eqref{eq:q0}, $b_0 = {M_\textrm{s}}/{Q_0}$. Combining this with the model limitation condition~\eqref{eq:rplus-gg-QQ0} (more precisely $M\gg Q_0$), one can see that $b_0 \gg 1$.
    Also, for reasons of numerical stability which are discussed in~\ref{S:results}, we rescale the time as $\tau\defi t\; s_0/M_\textrm{s}$ (we stress that $\tau$ is a rescaled time and not proper time). With this, the new evolution equations become:
        \begin{equation}\label{eq:dmudtHWnew}
		\frac{\dif \mu}{\dif \tau} = -\frac{ \left(H(\mu,Y)+S(\mu,Y) Y^2\right)}{ (\sqrt{1-Y}+1)^4},
	\end{equation}
    \begin{equation}\label{eq:dYdtHWnew}
		\frac{\dif Y}{\dif \tau} = \frac{2 \left( H(\mu,Y) -S(\mu,Y) \left(1-Y+\sqrt{1-Y}\right)Y\right)Y}{\mu (\sqrt{1-Y}+1)^4},
	\end{equation}
      When choosing mass scale of $M_\textrm{s}=10^8 M_\odot$, and setting $\alpha=\num{0.26792}$ (zero massless neutrinos) for comparison with figure~2 of \cite{Hiscock:1990ex}, the fiducial values associated with the original HW analysis are $s_0=\num{5.3065e-97}$, $z_0=\num{0.53027}$, $b_0=\num{586.09}$.

    \subsection{Configuration space analysis}
    \label{S:config-space}
    In this section we will dissect all relevant aspects of the configuration space evolution $Y(\mu)$ of RN black holes. All of the qualitative results presented here are valid both in the case of standard electromagnetism as well as in the dark $U(1)$ model we will further explore in section~\ref{S:DarkE}. 
    
    \subsubsection{Configuration space solutions}
    \label{S:config-space-sol}
    From the $Y$ and $\mu$ evolution equations~\eqref{eq:dYdtHWnew} and~\eqref{eq:dmudtHWnew}, one can obtain a differential equation for the configuration space solutions $Y(\mu)$:
    \begin{equation}\label{eq:dYdmuHWnew}
		\frac{\dif Y}{\dif \mu} = \frac{2 S(\mu,Y) \left(\sqrt{1-Y}+1\right) Y^2}{\mu  \left(H(\mu,Y)+S(\mu,Y) Y^2\right)}-\frac{2 Y}{\mu }.
	\end{equation}
     From now on, we will assume that the variables $\mu$ and $Y$ are both normalized, such that the evolution of $(\mu,Y)$ is contained in the domain $(0,1)\times(0,1)$. 
     
     Given that $H(\mu,Y)$ and $S(\mu,Y)$ represent the functions associated with the Hawking and Schwinger evaporation terms, one may evaluate the system's behaviour in the near-extremal regime and when setting each of these terms to zero individually.     
     In each case, equation~\eqref{eq:dYdmuHWnew} simplifies considerably, as the resulting equations for all such phases do not explicitly depend on either function $H(\mu,Y)$ or $S(\mu,Y)$. In fact, one can obtain explicit simple analytic solutions in each case. 

     \paragraph{Near-extremal phase:} By taking the limit $Y\to 1$ in equation~\eqref{eq:Haw}, we have $H(\mu,Y) \to 0$. Applying this to equation~\eqref{eq:dYdmuHWnew}, we have:
     \begin{equation}
         \frac{\dif Y}{\dif \mu} \stackrel{Y\to 1}{=} \;\frac{2 \;S(\mu,Y)}{\mu  \;S(\mu,Y) }-\frac{2 }{\mu } = 0.
         \label{eq:Yconst}
     \end{equation}
     Therefore, an approximation for the limiting near-extremal case is simply given by $Y = \textrm{const}$. This result was already expected, since in this regime the black hole's temperature drops significantly, and the system enters a \enquote{freeze-out} state, as can be seen in figure~\ref{fig:evaporation2b}.

\begin{figure}[!ht] 
		\begin{subfigure}{0.48\textwidth}
			\includegraphics[width=\linewidth]{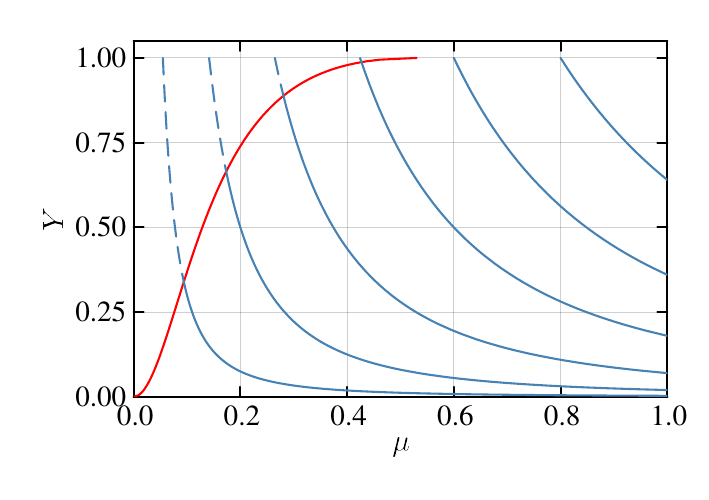}
		\end{subfigure}\hspace*{\fill}
		\begin{subfigure}{0.48\textwidth}
			\includegraphics[width=\linewidth]{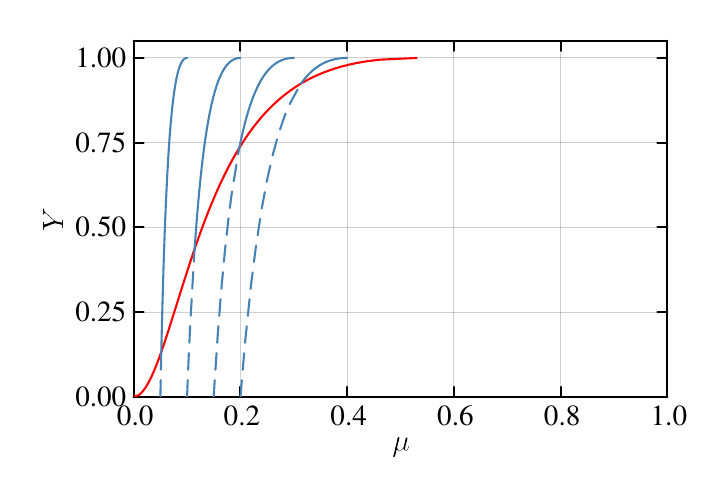}
		\end{subfigure}
  \caption{Hawking (left) and Schwinger (right) configuration space curves are presented in blue. The approximate attractor curve is depicted in red for $z_0=\num{0.53027}$.} \label{fig:hsanalytic}
    \end{figure}

     \paragraph{Hawking-dominated phase:} The Hawking-dominated phase corresponds to setting $S(\mu,Y)=0$. In this case, one has the solution:
    \begin{equation}\label{eq:YmuH}
        Y_\textrm{H}(\mu)=\left(\frac{\mu_{\textrm{h}}}{\mu}\right)^2,
	\end{equation}
    where $\mu_{\textrm{h}}$ is defined so that $Y_\textrm{H}(\mu=\mu_{\textrm{h}})=1$. The behaviour of these curves is presented in figure~\ref{fig:hsanalytic} (left). Note, however, that this is an approximation and no black hole ever really reaches $\mu_{\textrm{h}}$, since this is the point of extremality. Furthermore, depending on the initial $\mu$ and $Y$, the black hole may never even approach $\mu_{\textrm{h}}$ at all, since it intersects the attractor curve (red curve) beforehand. Such cases are represented by the partially dashed lines in the figure and equation~\eqref{eq:YmuH} is only expected to be valid below the attractor. For black holes that do reach the near-extremal limit, we call $\mu_{\textrm{h}}$ the \enquote{hanging mass} --- and, as we shall see in the numerical solutions, the system spends a significant portion of its lifetime near this mass. 
    Comparing the curves in figure~\ref{fig:hsanalytic} (left) with those in the mass dissipation zone in figure~\ref{fig:configuration-space}, we see that the regime dominated by Hawking evaporation $(S(\mu,Y) \sim 0)$ corresponds to what we previously called \enquote{mass dissipation zone}. 
    
    \paragraph{Schwinger-dominated phase:} The Schwinger-dominated phase corresponds to setting $H(\mu,Y)=0$. The solution is then given by:
    \begin{equation}\label{eq:YmuS}
		Y_\textrm{S}(\mu)=\frac{(2 \mu -\mu_1)\mu_1}{\mu ^2},
	\end{equation}
    where $\mu_1 = \mu(Y=1)$ and $\mu \leq \mu_1$. These curves are presented in figure~\ref{fig:hsanalytic} (right). Again, note that equation~\eqref{eq:YmuS} is only expected to be valid until the point where their trajectory $Y(\mu)$ meets the attractor curve, meaning the continuation of the dashed lines in figure~\ref{fig:hsanalytic} (right) must be discarded. Similarly, by comparing the curves in figure~\ref{fig:hsanalytic} (right) with those in the charge dissipation zone in figure~\ref{fig:configuration-space}, it is clear that the regime dominated by the Schwinger effect $(H(\mu,Y) \sim 0)$ corresponds to what we previously called \enquote{charge dissipation zone}. 

    \vspace{0.5cm}
    
\begin{figure}[t]
	\centering
		\begin{subfigure}{0.48\textwidth}
			\includegraphics[width=\linewidth]{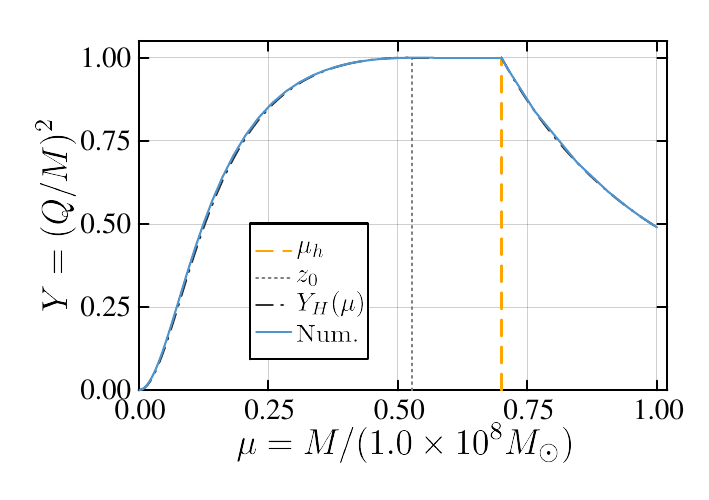}
		\end{subfigure}
        \begin{subfigure}{0.48\textwidth}
			\includegraphics[width=\linewidth]{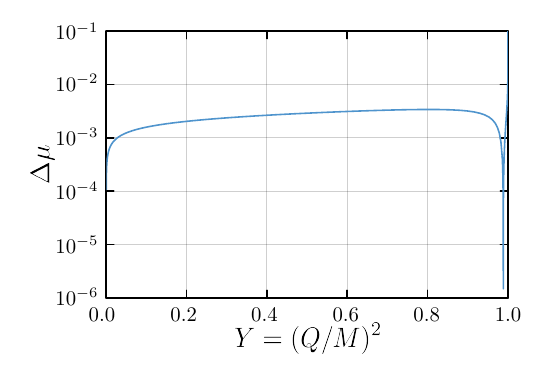}
		\end{subfigure}
	\caption{Configuration space solution for the parameter choices $z_0=\num{0.52682}$, $\mu_{\textrm{h}}=\num{0.7}$. On the left, we present a numerical solution superimposed on a piecewise solution constructed from the approximate analytic formulas. On the right, we plot the residual $\Delta \mu\defi |\mu_\textrm{analytical}-\mu_\textrm{numerical}|$ with respect to $Y=(Q/M)^2$. The large residual near $Y=1$ is due to the fact that the numerical solution does not reach $Y=1$, while the approximate attractor curve reaches $Y=1$ at $\mu=z_0$.} 
    \label{fig:pssanalytic}
\end{figure}

    The statements made above identifying the approximate curves in equations~\eqref{eq:YmuH} and~\eqref{eq:YmuS} to the evolution in the respective mass and charge dissipation zones of course go beyond the visual confirmation. We have also performed a comparison with the full numerical analysis and have seen that, as long as one does not approach the attractor region too closely, the approximate solutions perform very well as an approximation to the full solution, as we show in figure~\ref{fig:pssanalytic}.
    The Schwinger-associated function $S(\mu,Y)$ has an exponential form,
    so that the value of $S(\mu,Y)$ can be suppressed by a large negative exponent.
    Combined with the large overall multiplying factor $b_0$, 
    the sign of the function inside the exponential in $S(\mu,Y)$ will be crucial in defining on which evaporation zone a particular black hole initial state will be located. Using equation~\eqref{eq:S-eq}, we may (approximately) define the mass and charge dissipation zones as:
    \begin{equation}
        z_0-\frac{\mu \left(\sqrt{1-Y}+1\right)^2}{\sqrt{Y}} \qquad \left\{ \begin{array}{llc}
        >0 \quad\implies\quad \textrm{Charge dissipation zone}\\
        <0 \quad\implies\quad \textrm{Mass dissipation zone}
\end{array}\right.	\label{eq:zones}
    \end{equation}

    \subsubsection{Approximate attractor curve}
    \label{S:approx-attractor}
    Solutions from both the mass dissipation zone as well as from the charge dissipation zone flow towards the attractor region, {a basin of attraction in the configuration space $(\mu,Y)$ where the curves accumulate. We observe that in the attractor region,} the value of $S(\mu,Y)$ {must be} roughly of the same order as that of $H(\mu,Y)$ {(which in turn is roughly of order unity)}. In that region, the approximations used to obtain the solutions~\eqref{eq:YmuH} and~\eqref{eq:YmuS} fail.  Again, given the strong dependency on the $S(\mu,Y)$ exponential factor, this region can be characterized by the point at which the exponent in $S(\mu,Y)$ vanishes {(equivalently $S(\mu,Y)\sim1$)}, expressed as the condition:
    \begin{equation}
        \mu=z_0\frac{\sqrt{Y}}{(\sqrt{1-Y}+1)^2}.
        \label{eq:muatt}
    \end{equation}
Note that, since $Y\in(0,1)$ and $\mu(Y)$ is monotonic, the maximum value of $\mu$ on the approximate attractor curve is $\mu \to z_0$ as $Y\to 1$. 
Returning to the \enquote{hanging mass} concept, one can now see that \emph{whenever the black hole reaches the near-extremality, $\mu_{\textrm{h}} \geq z_0$}. This is an extremely important result which we will return to several times for the remainder of this paper.

Again, it is noteworthy that the true attactor never actually reaches the point of extremality --- it only asymptotically approaches it. Furthermore, the approximate attractor curve does not propagate, as the true attractor, to $\mu=1$ (it stops at $\mu = z_0$) as shown in figure~\ref{fig:hsanalytic}.  Nevertheless, this is a still useful expression, specially if one is interested in investigating the attractor position for different scenarios of dark electromagnetism, as we are in this paper.     
It should be also mentioned that an alternative definition for an approximate attractor may be found in \cite{Ong:2019vnv}, which also satisfies the rough description provided in \cite{Hiscock:1990ex}, namely that the attractor lies in the region where the rates of mass loss and charge loss are the same order of magnitude. However, the expression equation~\eqref{eq:muatt} is useful, as it provides for the first time a closed form analytic expression for an approximate attractor curve in the configuration space $(\mu,Y)$ for $\mu < z_0$ values.

One technical point which must be clarified now regards the dependence of the precise location of the end of the approximate attractor curve $(M_{z_0} : = z_0 M_\textrm{s})$ on the chosen mass scale normalization. One might recall that when introducing the modified evolution equations in section 
\ref{S:modified}, a mass scale $M_\textrm{s}$ was introduced in order to maintain the normalized mass $\mu$ in the interval $(0,1)$. By varying such a scale, the physical mass $M = \mu M_\textrm{s}$ obviously remains unchanged. The scale dependent parameters $\mu$ and $z_0$, however, must be properly readjusted. A small caveat comes from the fact that, while $\mu$ has a simple linear dependence on the mass scale, the dependence on $z_0$ is given by a nontrivial logarithmic relation (see equation~\eqref{eq:ConstDefs}) --- which scales as $M_{z_0} = z_0 M_\textrm{s} \sim O(1)+\ln(M_\textrm{s}^2)$. 
In terms of the \enquote{hanging mass} concept,  we see that black holes which do become near-extremal must satisfy $M_\textrm{s}\; \mu_\textrm{h} \geq M_{z_0}$. 
Fortunately, this dependence is rather weak. Given a reference mass scale of $M_\textrm{s}=10^8 M_{\odot}$ and a rescaling $M_\textrm{s}$ by factor $\sigma_{M}$ so that $M_\textrm{s}\rightarrow \sigma_{M} M_\textrm{s}$, one finds that $M_\textrm{s} z_0\rightarrow (1+\num{0.00644} \ln(\sigma_{M})) M_\textrm{s} z_0$. Applying this to the relevant range of scales considered in figure~\ref{fig:PBH4} --- which spans 40 orders of magnitude in mass --- the quantity $z_0 M_\textrm{s}$ is rescaled from its original value at $M_\textrm{s}=10^8 M_{\odot}$ by a factor between $\sim \num{0.6}$ and $\sim \num{1.2}$. In a similar manner, $z_0$ is weakly dependent on $\alpha$; rescaling $\alpha$ by a factor of $10$ yields less than a $1\%$ change in $z_0$.

Of course, the \enquote{true} attractor curve is ultimately independent of $M_\textrm{s}$, and the nontrivial dependence of the approximate attractor curve on $M_\textrm{s}$ ultimately follows from the assumption that $S(\mu,Y)\sim 1$ in the attractor region. 
Later, in section~\ref{S:rescaling}, we will identify rescaling relations that preserve the position of $z_0$ in configuration space. This will aid in mitigating possible ambiguities which might arise from the choice of the mass scale.
This issue, however, is rather technical and, as long as one carefully chooses a proper mass scale for one's given problem from the start, one shouldn't have to worry about rescaling effects at all in the numerics. Nevertheless, since the evaporation time estimates depend non-trivially on the mass scale and given that we will consider a relatively large range of PBH mass scales in our analysis (when moving to the scenario of dark electromagnetism), we will find it necessary to revisit the rescaling behavior when performing evaporation timescale estimates.

\subsubsection{Configuration space evolution}

Let us consider a black hole initially in the mass dissipation zone. When sufficiently far away from the attractor region, the configuration space evolution of the system in the mass dissipation zone is well-approximated by equation~\eqref{eq:YmuH}.
If $\mu_{\textrm{h}} <z_0$ (blue curves with dashed segments in figure~\ref{fig:hsanalytic} (left)), the system will evolve within the mass dissipation zone until it intersects the attractor curve. In this case the system never achieves the near-extremal limit. The same happens for a black hole starting from the charge dissipation zone --- it evolves along the curves approximated by~\eqref{eq:YmuS} until it reaches the attractor. Once in the attractor region, the evolution stays in this regime. The trajectory in the configuration space can then be described by means of the approximate attractor curve~\eqref{eq:muatt}.

If $\mu_{\textrm{h}}>z_0$, the system will evolve until it approaches the extremal curve $Y=1$. Once in the near-extremal regime, the system will evolve along a (nearly horizontal) trajectory on the configuration space (see equation~\eqref{eq:Yconst}) close to the extremal curve $Y=1$ until $\mu\approx z_0$.
At this point, the system will start evolving in a neighborhood of the attractor curve \cite{Hiscock:1990ex} losing both mass and charge in a similar rate. An illustration of the configuration space trajectory for this case is provided in figure~\ref{fig:pssanalytic}, where the mass dissipation zone $(\mu>\mu_\textrm{h})$, near-extremal regime $(z_0<\mu<\mu_\textrm{h})$ and attractor region $(\mu<z_0)$ are depicted.

\subsection{Evaporation time estimates}
\label{S:evap_time_est}

\subsubsection{Near-extremal solution}\label{S:extremal}
    
    Equations~\eqref{eq:dmudtHWnew} and~\eqref{eq:dYdtHWnew} admit a simple solution at the extremal limit $Y\to1$, which will be of interest later on. In this limit, ${\dif Y}/{\dif \tau} = 0$ (in agreement with equation~\eqref{eq:Yconst}) and ${\dif \mu}/{\dif \tau}$ simplifies to:
    \begin{equation}\label{eq:dmudtHWnewX}
		\frac{\dif \mu}{\dif \tau} = -  \exp[(z_0-\mu)b_0].
	\end{equation}
    This equation can then be integrated between near extremal points with different masses in order to obtain an expression for the evaporation time in terms of the respective initial and final mass parameters $\mu_\textrm{i}$ and $\mu_\textrm{f}$ of the black hole:
    \begin{equation}\label{eq:dTgen}
		\Delta \tau = -\left. \frac{\exp[(\mu-z_0)b_0]}{b_0}\right|^{\mu_\textrm{f}}_{\mu_\textrm{i}}.
	\end{equation}
    One may simplify this formula even further by recognizing that if $(\mu_{\textrm{h}} - z_0)\gg1/b_0$ and $b_0$ is of order unity or higher, the nontrivial exponential term dominates. Thereby we obtain the following expression (where $\Delta t \defi \Delta\tau\; M_\textrm{s}/s_0$):
    \begin{equation}\label{eq:dTgenXseg1}
		\Delta t 
        \sim \frac{M_\textrm{s}}{s_0 \;b_0} \exp[(\mu_{\textrm{h}}-z_0) \;b_0] 
	\end{equation}
    Later, we shall verify numerically that in the cases of interest, the system will spend a significant time near the extremal limit $Y=1$, and for solutions in which $\mu$ varies significantly, one can obtain an order of magnitude estimate for the \emph{full} evaporation time from the near-extremal segment of the solution. 
    It is worth reminding ourselves that extremality is never achieved, but only asymptotically approached in some specific $Y$ and $\mu$ ranges \cite{Ong:2019vnv}.

\subsubsection{\texorpdfstring{Small $Y$ limit}{Small Y limit}}
\label{S:small-y}

In order to obtain the evaporation time in the limiting case of small $Y$, let us start from equation~\eqref{eq:dmudtHWnew}:
\begin{equation}
	\dif \tau = -\frac{(\sqrt{1-Y}+1)^4 }{ \left(H(\mu,Y)+ S(\mu, Y) Y^2\right)}\dif \mu,\label{eq:dtau}
\end{equation}
where $H(\mu,Y)$ and $S(\mu,Y)$ are given by equations~\eqref{eq:Haw} and~\eqref{eq:S-eq}, respectively. By taking the $Y\to 0$ limit, we obtain:
\begin{equation}
    \stackrel{Y \to 0}{\lim} \dif \tau = -\frac{32}{27}\mu^2 \dif\mu\;.
    \label{eq:schwarzschild_lim}
\end{equation}
Integrating, we have:
\begin{equation}
    \Delta \tau \;\stackrel{Y\to0}{=}\; -\int_{\mu}^{0}\frac{32}{27}\tilde\mu^2 \dif\tilde\mu\;\quad\to\quad \Delta\tau_\textrm{schw} \;\stackrel{Y\to0}{=}\; \frac{32}{81}\mu^{3}\;.\label{eq:dtau_Y=0lim}
\end{equation}
Therefore, as expected, once we are in the small $Y$ regime, the evaporation time result for the Schwarzschild solution is recovered. This solution is valid regardless of the black hole's location in the configuration space (near the attractor or within the charge and mass dissipation zones). Since this result will be useful for future sections, let us look more closely at the small $Y$ behaviour in the mass dissipation zone and attractor region.

\paragraph{Mass dissipation zone}
Taking $S(\mu,Y)\to0$ in equation~\eqref{eq:dtau} and Taylor expanding $\dif\tau$ up to second order in $Y$, we have:
\begin{equation}
    \dif\tau = \left[ -\frac{32\;\mu^2}{27} -\frac{32\;\mu^2}{81}Y -\frac{344}{729}\mu^2\;Y^2 \right]\dif\mu\;.
    \label{eq:smallY_time_MDZ}
\end{equation}
For small $Y$, the first two terms will be dominant.

\paragraph{Attractor region}
Taking $S(\mu,Y)\to1$ in equation~\eqref{eq:dtau} and Taylor expanding $\dif\tau$ up to second order in $Y$, we have:
\begin{equation}
    \dif\tau = \left[ -\frac{32\;\mu^2}{27} -\frac{32\;\mu^2}{81}Y -  \left(\frac{344}{729} -\frac{64\;\mu^2}{729}\right)\mu^2\;Y^2\right]\dif\mu\;.
    \label{eq:smallY_time_attr}
\end{equation}
As before, the first two terms are the dominant terms in the small $Y$ regime.

Equations~\eqref{eq:smallY_time_MDZ} and~\eqref{eq:smallY_time_attr} show that the difference between the evaporation times in the two regions only appears in the second order term in $Y$. The attractor region evaporation time contains an extra positive term proportional to $\mu^4Y^2$. Note, however, that in the attractor region, taking the small $Y$ limit simultaneously imposes a small $\mu$ limit, suppressing even further the time difference between the two regimes. This result will be further discussed and applied in section~\ref{S:Muniv}, where we will obtain updated minimum bounds for dark-charged PBHs to live longer than the age of the universe.

\subsubsection{\texorpdfstring{Attractor timescale estimate for $\mu\leq z_0$}{Attractor timescale estimate for μ<z₀}}\label{sec:timeest}

\begin{figure}[t]
	\centering
    \begin{subfigure}{0.48\textwidth}
		\includegraphics[width=\linewidth]{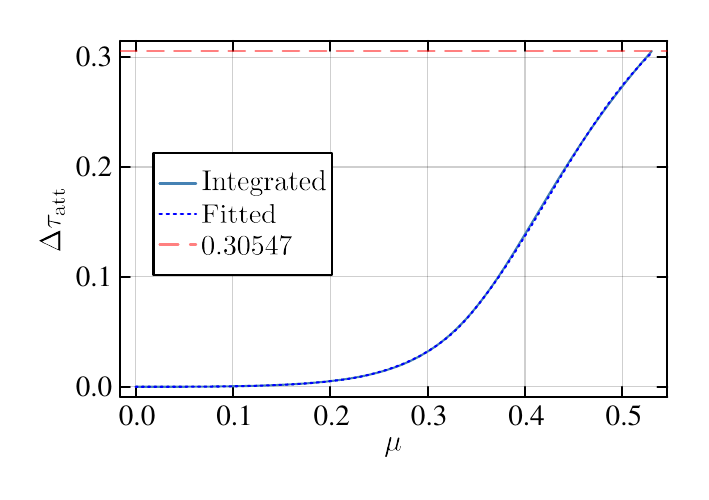}
	\end{subfigure}
    \begin{subfigure}{0.485\textwidth}
		\includegraphics[width=\linewidth]{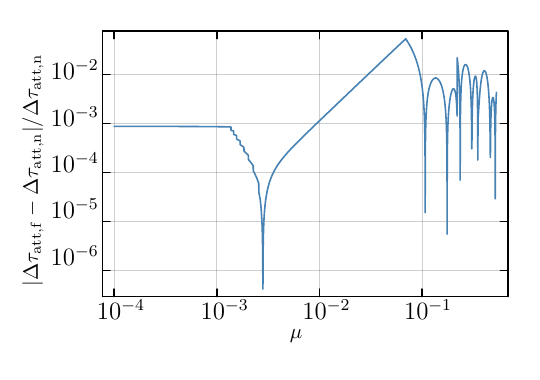}
	\end{subfigure}
	\caption{Plots showing the agreement between the rescaled evaporation time along the attractor region. On the left we show the analytic result for the approximate attractor curve given by equation~\eqref{eq:dtaumuattr1} (Integrated) together with a fitted function constructed with equation~\eqref{eq:dtau(mu)} (Fitted). The solution for $\Delta\tau_{\mathrm{att}}$ was obtained using the value $z_0=\num{0.53027}$ for standard electromagnetism (setting $\alpha=0.26792$, following figure~\ref{fig:hsanalytic}). 
    The horizontal line near the top has the value $\Delta\tau_{\mathrm{att}}(\mu=z_0)=\num{0.30547}$, which corresponds to the total time the system spends between $\mu=z_0$ and $\mu=0$. On the right we present a plot of the residual between the fit and the integral; for the most part, the fit differs by less than one part in a hundred.} 
    \label{fig:tauattranalytic}
\end{figure}

Given the attractor curve formula~\eqref{eq:muatt}, one can obtain a time estimate for the time spent on the attractor curve by direct integration. From equation~\eqref{eq:dmudtHWnew}, we obtain:
\begin{equation}\label{eq:dtaumuattr0}
	\dif \tau = -\frac{(\sqrt{1-Y}+1)^4 \dif \mu}{ \left(H(\mu,Y)+ Y^2\right)},
\end{equation}
where we make use of the fact that $S(\mu,Y)=1$ when the condition~\eqref{eq:muatt} holds. Since equation~\eqref{eq:muatt} is difficult to invert, we perform the following variable transformation (with $x\in(0,1)$):
\begin{equation}\label{eq:muattrpar}
	\mu = \frac{1}{2} \sqrt{x} (x+1) z_0,
    \qquad
    Y =\frac{4 x}{(x+1)^2}.
\end{equation}
Applying this variable change to equation~\eqref{eq:dtaumuattr0} yields the integral expression:
\begin{equation}\label{eq:dtaumuattr1}
	\Delta \tau_{\mathrm{att}}(x) = \int_0^x \frac{4  z_0 (3 x'+1)}{\sqrt{x'}} \left[\frac{[(1-x')P(x')]^4}{4 x' z_0^2 [8 x'-(x'+1)P(x')]}-16 x'^2\right]^{-1} dx',
\end{equation}
where $P(x):=\sqrt{9 \left(x^2+1\right)-14 x}+3 (x+1)$. One can verify (by Taylor expansion) that for small $x$, one recovers equation~\eqref{eq:dtau_Y=0lim}. The function $\Delta \tau_{\mathrm{att}}(x)$ provides an estimate for the (rescaled) time the system spends between $\mu=0$ and $\mu=\mu(x)\leq z_0$. The function $\Delta \tau_{\mathrm{att}}(x)$ is parametrically plotted with $\mu(x)$ in figure~\ref{fig:tauattranalytic}. The dimensionful time estimate is $\Delta t_{\mathrm{att}}:=\Delta \tau_{\mathrm{att}}M_\textrm{s}/s_0^\chi$.

Since we are mainly interested in the order of magnitude of the evaporation time (along the attractor curve), we have proceeded by fitting\footnote{We perform a least-squares fit using the Julia language package \texttt{LsqFit.jl}.} piecewise auxiliary functions to the parametric curve $\Delta \tau_{\mathrm{att}}(\mu)$ generated from equations~\eqref{eq:muattrpar} and~\eqref{eq:dtaumuattr1}. Also, given that in the following sections we will be interested in obtaining the inverse function $\mu(\Delta\tau_{\mathrm{att}})$, we have chosen auxiliary functions which are easily invertible. The resulting piecewise function is given by:
\begin{equation}
\label{eq:dtau(mu)}
    \Delta \tau_{\mathrm{att}}(\mu) = \frac{1}{a + b\;\mu^{-m}}\;, 
\end{equation}
where in the small $\mu$ regime $(\mu < \num{0.07})$ we have used the analytic expression~\eqref{eq:dtau_Y=0lim}, which corresponds with setting the parameters $(a,b,m)$ to $(0, 81/32, 3)$. 
For the low range $(\num{0.07} \leq \mu < \num{0.22})$, the best fit parameters obtained are $(a,b,m) = (-\num{42.78}, \num{1.258}, \num{3.264})$, for the medium range $(\num{0.22} \leq \mu < \num{0.35})$, the best fit parameters are $(a,b,m) = (-\num{3.848}, \num{0.165}, \num{4.451})$, and for the upper range $(\num{0.38} \leq \mu < z_0=\num{0.53027})$, the best fit parameters are $(a,b,m) = (\num{2.490}, \num{0.01351}, \num{6.411})$.
A plot showing the agreement between the best fit piecewise function and $\Delta\tau_{\mathrm{att}}$ can be seen in figure~\ref{fig:tauattranalytic} (left). Note that equation~\eqref{eq:dtau(mu)} is only valid for estimating the rescaled evaporation time along the approximate attractor curve and therefore should only be applied to values of $\mu \leq z_0$. Besides being an approximation, the agreement between the curves is very good for the values obtained from integrating equation~\eqref{eq:dtaumuattr1}, as can be seen in the plot of the residuals shown in~\ref{fig:tauattranalytic} (right).
The significance of these results will be further discussed in section~\ref{S:near_extr_time}. There we will implement rescaling transformations which will allow us to obtain the evaporation times spent near the attractor curve for Reissner--Nordström PBHs in different scenarios of dark electromagnetism.

    \section{Dark-charged PBHs}\label{S:DarkE}

    In this work, we adopt a simple $U(1)$ dark electromagnetism model, with a massless dark photon and a single massive charged dark particle $\chi$ --- the dark electron. In the case of $m_\chi = m_e$ and $e_\chi = e$, this model has been refereed to as the \enquote{mirror sector}. In here, we will allow for these parameters to vary. For further discussion of symmetry broken mirror dark matter models see \cite{Foot:2014mia,Mohapatra:2001sx} and references therein.  
    Following along the line of~\cite{Ackerman:2008kmp} and \cite{Agrawal:2016quu}, the Lagrangian for the system is given by:
    \begin{equation}
        \mathcal{L} = \Bar{\chi}(i\slashed{D} + m_{\chi})\chi -\frac{1}{4} {F}_{\mu\nu}{F}^{\mu\nu}\;.
        \label{eq:lagrangian}
    \end{equation}
    Here ${F}_{\mu\nu}$ is the field-strength tensor for dark electromagnetism, $D_{\mu} = \partial_\mu -i e_{\chi} {A}_\mu$ is the gauge covariant derivative, $\slashed{D}$ is the dark Dirac operator,  and $m_\chi$ and $e_{\chi}$ are the dark electron mass and charge respectively. 
    
    Note that this is the simplest possible dark $U(1)$ model, without any coupling whatsoever between the dark particles and standard model particles. In the analysis that follows, allowing for the presence of couplings could possibly significantly vary the dynamics and evolution of the dark-charged PBHs. Although understanding the effects of having a more \enquote{sophisticated} model of dark electromagnetism model is of high importance, we have decided to focus solely on how the presence of a dark $U(1)$ charge impacts the PBHs evolution, leaving more realistic models for future work.

    We will be considering, from now, Reissner--Nordström black holes charged from such a $U(1)$ field. The aim is to investigate the effects coming from varying the dark electron mass and charge in the attractor's location, evaporation time-scales, and investigate which regions of the $(m_\chi, e_\chi, M_\textrm{PBH})$ parameter space are still feasible as dark matter candidates.

    \subsection{\texorpdfstring{$(m_{\chi}, e_{\chi}, M_{\normalfont\textrm{PBH}})$ parameter space}{(mχ, eχ, MPBH) parameter space}}\label{s:parameter space}

    \begin{figure}[!ht]
		\begin{subfigure}{0.48\textwidth}
		\includegraphics[width=\linewidth]{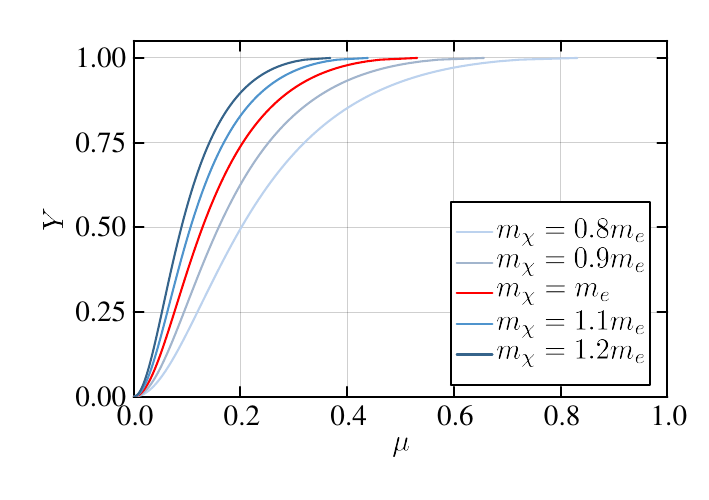}
		\end{subfigure}\hspace*{\fill}
		\begin{subfigure}{0.48\textwidth}
		\includegraphics[width=\linewidth]{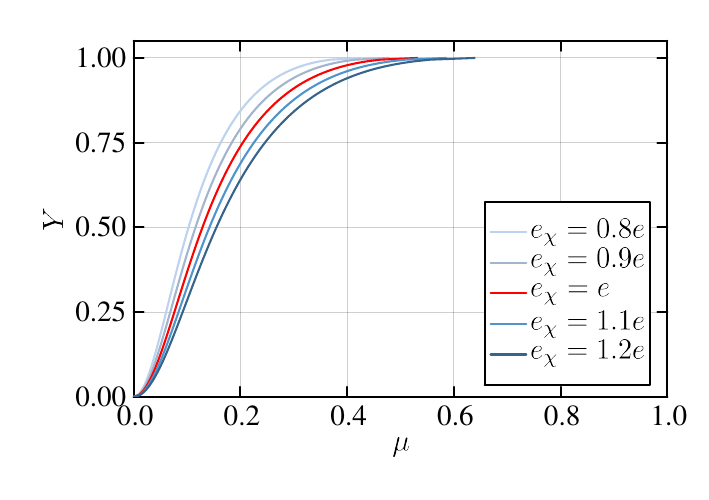}
		\end{subfigure}
  \caption{Attractor curves~\eqref{eq:muatt} for various dark electron masses and charges.}
        \label{fig:DE_attractors}
	\end{figure}
 
    Constraints derived in the literature for the $U(1)$ dark electromagnetism model here adopted were discussed in detail in~\cite{Ackerman:2008kmp, Feng:2009mn, Agrawal:2016quu}.
    The constraints come from both the evolution of dark matter in the early universe as well as from observed galactic dynamics. In summary, a feasible dark matter model must provide~\cite{Ackerman:2008kmp, Agrawal:2016quu}:
    \begin{itemize}
        \item the correct relic density given by the chosen freeze-out mechanism;
        \item the correct ellipticity bounds on galaxy halos;
        \item the correct rate of dwarf galaxy survival as they orbit around the host galaxy;
    \end{itemize}
    among others. These allow obtaining bounds on the dark electron mass $m_{\chi}$ and dark fine-structure constant $\alpha_\chi$. In order to give an idea on the orders of magnitude, masses $m_{\chi}$ on the GeV scale are in a safe allowed zone for fine structure constants $\alpha_{\chi}\lesssim 10^{-4}$, while masses $m_{\chi}$ in TeV scales are allowed for $\alpha_{\chi}\lesssim 10^{-2}$ (see Fig 3 in \cite{Ackerman:2008kmp} and Fig 4 in \cite{Agrawal:2016quu} for further details).
    
    It is worth mentioning that the most restrictive constraint comes from the relic abundance at thermal freeze-out, which might rule out significant regions in the allowed $(m_\chi, e_\chi)$ parameter space, as it is indeed the case in \cite{Ackerman:2008kmp}.     
    Note, however, that  as discussed in section~\ref{S:Introduction}, we do not really need or even expect these dark particles to still be present in the current epoch of our universe. The dark matter model proposed herein is simply the dark-charged PBHs themselves. And, as we shall see in the following sections, the necessary mass the dark electron must have in order to extend the lifetime of extremely light PBHs beyond the age of the universe is of the order of a \si{\tera\electronvolt} (higher or lower depending on the dark electron charge chosen). Given the mass range of $m_\chi$, it is therefore reasonable to expect them to exist only during the primordial, hot, dense stages of the universe, as long as they exist long enough for the formation of  dark-charged PBHs. Afterwards, the necessary relic abundance at thermal freeze-out can be virtually nil. Therefore, it is clear that most of the constraints present in the literature do not directly apply to our case.
    
    One might argue, of course, that once these dark-charged PBHs start to evaporate and lose charge, dark matter particles will be emitted, making the soft and hard scattering constraints relevant again. This is an important point, which we will leave for further study in the future. 
    As already extensively discussed, the initial PBH's charge-to-mass ratio required to extend its lifetime (and even achieve near-extremal regimes) can start significantly low, depending on where in the configuration space a given PBH is located. This implies that, when calculating such bounds, one will need to explore all the degrees of freedom this problem imposes with care. 
    As a quick example, in order to grasp the orders of magnitude involved, let us consider a PBH with initial mass of $10^{-19}M_\odot$ and an initial charge-to-mass ratio of $Y=10^{-2}$.  Then assuming $e_\chi=10^{-4}e$ and $m_\chi = 10^{11} m_e$ (values taken from figure~\ref{fig:DE_mass_charge}), the total mass emitted in dark particles --- in case this black hole actually fully evaporates, which may not even be the case --- would be about $4\times10^{-28}M_\odot\; (\sim \SI{1e3}{\kilo\gram})$\footnote{Note that this mass is precisely $10^{15}$ times the mass which would be emitted in electrons in the case of standard electromagnetism. This is as expected, since $\sigma_m/\sigma_e = (m_\chi/m_e)/(e_\chi/e) = 10^{15}$.} during a time of roughly $10^{15}$ years (timescale obtained from figure~\ref{fig:DE_mass_charge_time}). 
    Further studies analyzing combined scenarios of (uncharged) PBHs and WIMPs can also be found in the literature  \cite{Kadota:2021jhg, Gines:2022qzy, Oguri:2022fir}. Again, confronting the specific model here presented against dark matter constraints is left for future work.

    Taking the scenario described above as a valid assumption, let us now understand how varying the parameters $(m_\chi, e_\chi)$ affects the results for the evolution of charge black holes obtained so far. Keeping in mind that the attractor's position can be roughly defined by the parameter $z_0$ (see equation~\eqref{eq:ConstDefs}), this can be generalized to the dark sector as:
    \begin{equation}\label{Eq:z0chi}
        z^\chi_0=\frac{e_\chi \hbar}{\pi m_\chi^2 M_\textrm{s}}\ln\left(\frac{960 e_\chi^4 M_\textrm{s}^2}{\pi ^2 \alpha m_\chi^2 \hbar ^2}\right)\;.
    \end{equation}
    We can then see that, for higher dark electron masses and for lower $e_\chi$ charges, the value of $z_0^\chi$ decreases. In terms of the attractor's position in the configuration space, this leads to a shift to lower $M_\textrm{PBH}$ masses. This behaviour is illustrated in figure~\ref{fig:DE_attractors}, where one can see the effect on the attractor's position by varying the dark electron mass $m_\chi$ (on the right) and charge $e_\chi$ (left panel).
    Figure~\ref{fig:DE_attractors}, together with equation~\eqref{Eq:z0chi}, also reveals that the dependence of the attractor's position on $m_\chi$ and $e_\chi$ is not the same, being more sensitive to changes in mass than in charge, as one might expect from their appearance in the Schwinger effect.
    This clarifies prior work \cite{Ong:2019rnn}, which somewhat obfuscated these issues by focussing primarily on the charge-to-mass ratio $e_\chi/m_\chi$.

    Now, given that we are interested in low mass PBH, with
    masses lower than $\lesssim 10^{-15}M_{\odot}$, 
    a useful tool is to investigate the $(m_\chi, e_\chi)$ parameter region which shifts the attractor up to those mass scales. 

    \subsection{Rescaling}
    \label{S:rescaling}
    One might observe that the evaporation equations $\dif\mu/\dif\tau$ and $\dif Y/\dif\tau$ (see~\eqref{eq:dmudtHWnew} and~\eqref{eq:dYdtHWnew}) depend only on the parameters $z_0$ and $b_0$ (we note that $s_0$ has been absorbed into the definition of $\tau$).
    In this way, by obtaining the solutions of $\Delta\tau$ for the case of standard electromagnetism, one can extend these results to the dark sector 
    by rescaling the electron mass $m_e$ and charge $e$, while adjusting the mass scale $M_\textrm{s}$ of the system.
    For this reason, it is perhaps appropriate to consider in detail 
    how the constants $z_0$, $b_0$, and $s_0$ behave under parameter rescalings.
    Also, given the strong dependency of $\Delta\tau_\textrm{att}(z_0)$ on $z_0$, we are ideally looking for rescaling relations which allow us to keep $z_0$ fixed.

    First of all, note that under a rescaling $M_\textrm{s}^\chi = \sigma_{M} M_\textrm{s}$, $e_\chi = \sigma_{e} e$, and $m_\chi = \sigma_{m} m_e$, equation~\eqref{eq:ConstDefs} yields the following transformations of $s_0$, $z_0$, and $b_0$:
    \begin{equation}\label{eq:ConstTransformGen}
        s_0^\chi = \frac{s_0}{\sigma_{M}^2} ,\qquad 
        z_0^\chi = \frac{\sigma_{e}}{\sigma_{m}^2 \sigma_{M}}\left[z_0+\frac{2}{b_0}\ln\left(\frac{\sigma_{e}^2 \sigma_{M}}{\sigma_{m}}\right)\right],\qquad
        b_0^\chi =  \frac{\sigma_{m}^2 \sigma_{M}}{\sigma_{e}} b_0.
	\end{equation}
    It is not too difficult to show that the constants $z_0$ and $b_0$ are invariant if $\sigma_{e}=\sigma_{m}=1/\sigma_{M}$, which corresponds to transformations that preserve the charge-to-mass ratio $\varrho=e/m_e$. 
    It follows that for a fixed charge-to-mass ratio $\varrho$, one solution of equations~\eqref{eq:dmudtHWnew} and~\eqref{eq:dYdtHWnew} corresponds to all solutions of the original system~\eqref{eq:dMdtHWfull} and~\eqref{eq:dQdtHWfull} that are related by the transformation $\sigma_{e}=\sigma_{m}=1/\sigma_{M}$. Note also that under a rescaling of $M_\textrm{s}$, the coordinate time $t$ is rescaled as $t \rightarrow t \sigma_{M}^3$ for fixed $\tau=t s_0/M_\textrm{s}$.
    
    We would also like to consider transformations that allow for changes in the charge-to-mass ratio while leaving $z_0$ unchanged so that $z^\chi_0=z_0$. Such transformations are useful, since they maintain the position of the approximate attractor curve~\eqref{eq:muatt} in configuration space, while allowing for additional freedom in changing the dark electromagnetism parameters. 
    One can parameterize such a transformation as:
    \begin{equation}\label{eq:ConstTransformPar}
        \sigma_{e} = \frac{\vartheta}{\sigma_{M} \xi}
        ,\qquad 
        \sigma_{m} = \frac{\vartheta}{\sigma_{M} \xi^2}
        ,\qquad
        \xi = \xi(\vartheta) = \left(\frac{b_0 z_0 \vartheta}{b_0 z_0+2 \ln\vartheta}\right)^{1/3}.
	\end{equation}
    where $\xi=\sigma_{e}/\sigma_{m}$ may be interpreted as the change in the charge-to-mass ratio $\varrho^\chi=\xi\varrho$, and $\vartheta \in (\exp(1 - b_0 z_0/2),\infty)$ is the transformation parameter (the lower limit corresponds to the minimum of $\xi$). 
    Under this transformation, 
    \begin{equation}
    z^\chi_0=z_0\;, \qquad b^\chi_0 = b_0 + \frac{2}{z_0}\ln\vartheta \;. 
    \label{eq:z0fix}
    \end{equation}
     That there is a lower limit to the rescaling $\xi$ of the charge-to-mass ratio should not be surprising; from equation~\eqref{Eq:z0chi}, one can see that the charge $e_\chi$ can be lowered to a finite value that makes $z_0$ vanish. Moreover, from equation~\eqref{eq:dQdtHWfull} (noting $Q_0 \propto e$), the error function term dominates in the limit of small charge (holding everything fixed); in the following section, we will show how a rescaling to arbitrarily small values of the charge-to-mass ratio violates condition (iii), therefore falling outside the valid parameter range for equations~\eqref{eq:dmudtHWnew} and~\eqref{eq:dYdtHWnew}.

    Following the discussion presented in section~\ref{S:approx-attractor}, note that these rescaling formulae can be used to obtain an expression for the mass $M_{z_0}= z_0 M_\textrm{s}$ as a function of the rescaling parameters. In particular, one can use equation~\eqref{eq:ConstTransformPar} to obtain $\sigma_{M}=\vartheta/(\sigma_{m} \xi^2)$, yielding the formula:
    \begin{equation}\label{eq:MzTransformPar}
        M_{z_0}^\chi = \frac{\vartheta^{1/3}}{\sigma_{m}} \left(\frac{b_0 z_0+2 \ln\vartheta}{b_0 z_0}\right)^{2/3} M_{z_0} .
	\end{equation}
    The above formula is useful, since it provides a concrete estimate for the mass scale associated with the parameter $z_0$ under a rescaling of the parameters.

    \subsection{Updated model assumptions for the dark electromagnetism case}
    \label{S:updated-cond}
    
        At the end of section~\ref{S:assumptions} we have discussed the assumptions and limitations in the HW model. Those were summarized in three conditions:
        \begin{equation}
        \textrm{(i)}\;\; M \gg \frac{\hbar}{m_e}, \qquad \textrm{(ii)} \;\;\frac{e^3 Q}{m_e^2 r^2} \ll 1, \quad \textrm{and} \quad \textrm{(iii)}\;\; r_+^2 \gg QQ_0\;.
        \tag{\ref{eq:summ_conditions}}
    \end{equation}
        Let us now update these conditions assuming that the dark electron parameters are given by $m_\chi = \sigma_{m} m_e$, $e_\chi = \sigma_{e} e$. This will give us the new validity bounds for the \enquote{dark-charge PBH} where our analysis is valid. In order to avoid confusion with the previous mass constraints, we will refer to the \enquote{dark-charge PBH} mass as $M_{\textrm{DE}}$ in the following.

    \paragraph{Condition (i)} gives us:
        \begin{equation}
            M_{\textrm{DE}} \gg  \frac{\hbar}{m_\chi} = \frac{10^{-15}}{\sigma_{m}}M_{\odot}\;,
            \label{condition1}
        \end{equation}
        where we have used equation~\eqref{eq:constr1}.
        Note that this validity constraint does not affect the allowed values of $\sigma_e$, only those of $\sigma_m$ and $M_\textrm{DE}$ (or $\sigma_M$). In section~\ref{S:Muniv}, we will explicitly discuss the region in parameter space where this condition plays a role.
     
     \paragraph{Condition (ii)} gives us:
     \begin{equation}
     \frac{e_\chi^3 Q}{m_\chi^2 r^2} \ll 1\;.
            \label{eq:condition2}
        \end{equation}
        In order for this condition to be valid in the entire domain of outer communication of the PBH, one may replace $r$ by $r_+$, implying that:
    \begin{equation}
     (M + \sqrt{M^2 - Q^2})^2  = M^2 (1+\sqrt{1-Y}) \gg \frac{e_\chi^3 Q}{m_\chi^2}\;.
        \end{equation}
        Using the fact that $M = \mu M_\textrm{s}$, we have:
\begin{equation}
     \frac{\mu M_\textrm{s}\;(1+\sqrt{1-Y})^2}{\sqrt{Y}} \gg \frac{e_\chi^3}{m_\chi^2}\;.
        \end{equation}
    Near the attractor, in view of~\eqref{eq:muatt}, we may rewrite this expression as:
        \begin{equation}
     M_\textrm{s} z_0 = M_{z_0} \gg \frac{e_\chi^3}{m_\chi^2}\;.
        \end{equation}
    In section~\ref{S:Muniv} we will also explicitly discuss the region of parameter space where this condition plays a role in our analysis.

     \paragraph{Condition (iii)} In order to extend this condition to the scenario of dark electromagnetism, note that it can be rewritten as
     \begin{equation}
         \frac{(M + \sqrt{M^2 - Q^2})^2}{QQ_0} \gg 1,
     \end{equation}
     implying that
     \begin{equation}
         \frac{\mu M_s\;(1 + \sqrt{1 - Y^2})^2}{Q_0 \;\sqrt{Y}}  = \frac{b_0\;\mu\;(1 + \sqrt{1 - Y^2})^2}{\sqrt{Y}}\gg 1\;.
     \end{equation}
     Here, we have used the fact that $Q_0 = M_s/b_0$. Recalling the definition of the approximate attractor curve given by equation~\eqref{eq:muatt}, we may then rewrite this as:  
     \begin{equation}
        z_0\; b_0\;\mu \gg \frac{z_0\;\sqrt{Y}}{(1 + \sqrt{1 - Y^2})^2} = \mu_{\textrm{att}}\;.
     \end{equation}
     Remembering that for the standard electromagnetism case we had $z_0=\num{0.53027}$, and $b_0=\num{586.09}$, this implies that condition (iii) can be rewritten as:    
     \begin{equation}
        \mu \gg \frac{\mu_{att}}{b_0z_0} \sim \frac{\mu_{att}}{\num{310,8}}\;.
        \label{eq:z0b0chi}
     \end{equation}
     Therefore, as long as one is concerned with evolution along the attractor region and/or the mass dissipation zone, this condition does not impose any additional restriction on the mass range being analyzed. This is not a surprising result, given that this condition followed from the Schwinger effect, which is of order unity in the attractor region and exponentially suppressed in the mass dissipation zone.
     
     In order to extend this relation to the scenario of dark electromagnetism, let us make use of the rescaling relations given by equation~\eqref{eq:z0fix}, giving us: 
     \begin{equation}
         z_0^\chi b_0^\chi= b_0 z_0 +2\ln\left( \frac{\sigma_e^2\sigma_M}{\sigma_m}\right)\;.
     \end{equation}
     Applying this to equation~\eqref{eq:z0b0chi}, we have:
     \begin{equation}
        \mu^\chi \gg \frac{\mu_{att}^\chi}{b_0^\chi z_0^\chi } = \frac{\mu_{att}^\chi}{b_0 z_0 +2\ln(\sigma_e^2\sigma_M/\sigma_m)}\;.
     \end{equation}
     This implies that as long as $\ln(\sigma_e^2\sigma_M/\sigma_m)\gg -b_0 z_0/2 \sim -\num{1.5e2}$, the validity of condition~(iii) will always be satisfied as long the PBH trajectory is maintained in the attractor region and/or mass dissipation zone. This implies:
     \begin{equation}
         \frac{\sigma_e^2\;\sigma_M}{\sigma_m}\gg 10^{-66}\;.
     \end{equation}
    In all of the analysis presented in this article, this limit is never approached in the slightest, meaning that condition~(iii) will play no role in constraining any of the results presented here. It is also worth mentioning that condition (iii) ensures that the rescaling limit for $\xi$ discussed after equation~\eqref{eq:z0fix}, in which $b_0^\chi\rightarrow0$, is never approached.

 \subsubsection{The Validity Regime for the Schwinger Pair Production Rate}
\label{S:Validity}
	
 Another condition which must checked is the validity of the Schwinger pair production rate given by equation~\eqref{eq:Schwinger} in the new parameter space $(m_\chi, e_\chi, M_\textrm{PBH})$. 	The key constraint to keep in mind is that electric field gradients must be small compared to the Compton wavelength of the lightest electrically charged particle, given that this will be the easiest charged particle to be produced.
    In the standard model of particle physics, the lightest electrically charged particle is the electron. In our model, this mass will be given by $m_\chi$ for the dark electron. In this case, we must demand:
	\begin{equation}
		\frac{1}{E} \; \frac{\dif E}{\dif \textrm{(proper~length)}} < \frac{1}{\lambda_{\textrm{compDE}}} = \frac{m_\chi}{\hbar c}.
	\end{equation}
	In a Reissner--Nordström spacetime, keeping the geometrodynamic units $G\to1$, $c\to 1$, and $\hbar \to m_{\textrm{Planck}}^2$, and recalling that $E=q/r^2$,  we have:
	\begin{equation}
		\frac{1}{E} \; \frac{\dif E}{\dif \textrm{(proper~length)}}
		= \frac{\dif r}{\dif \textrm{(proper~length)}} \; \frac{1}{E} \; \frac{\dif E}{\dif r} =
		\sqrt{1-\frac{2M}{r} +\frac{Q^2}{r^2}}\; \frac{2}{r}\;.
	\end{equation}
	So, for applicability of the Schwinger calculation we demand
	\begin{equation}
		\sqrt{1-\frac{2M}{r}+\frac{Q^2}{r^2}} \;\; \frac{2}{r} <  \frac{1}{\lambda_{\textrm{compDE}}} = \frac{m_\chi}{m_{\textrm{Planck}}^2}.
    \label{eq:Schwinger_cond}
	\end{equation}

  \begin{figure}
		\centering		\includegraphics[width=0.8\linewidth]{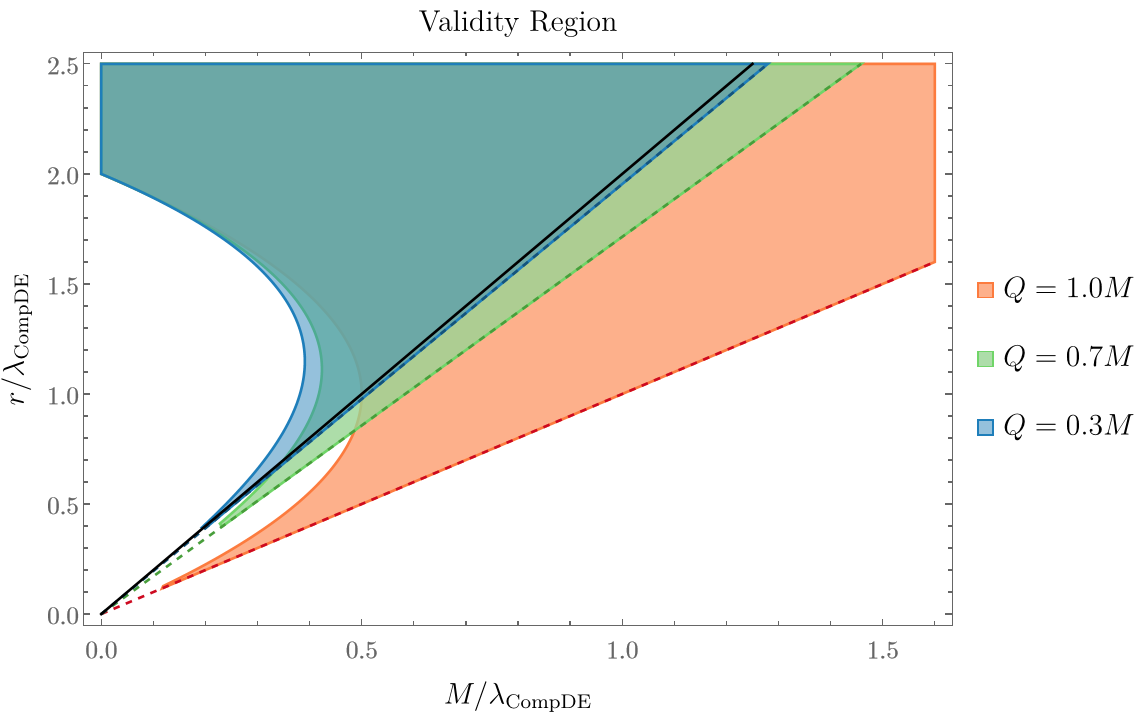}
		\caption{Plot showing the validity region of Schwinger's calculation outside of a charged black hole with different $Q/M$ ratios. Dashed lines represent the location of the event horizon for each case, the solid black line representing $Q=0$. Note that as one decreases the black hole's charge-to-mass ratio, the dashed lines start to 
        pile up near the black line for a horizon of zero charge.
        }
		\label{fig:validity-region}
	\end{figure}
 
	This condition is satisfied in two regions: at and close to the horizon and at large distances.
	We also need to demand that $r > r_+ = M + \sqrt{M^2-Q^2}$. 
	Squaring both sides of equation~\eqref{eq:Schwinger_cond}, we can re-write it as a quartic in $r$:
	\begin{equation}\label{eq:simult1}
		r^4 -4 \lambda_{\textrm{compDE}}^2 (r^2 -2Mr + Q^2 )  > 0\;,
	\end{equation}
    which we must simultaneously satisfy together with $r>r_+$ condition, that can be rewritten as:
	\begin{equation}
		r^2 - 2Mr + Q^2 >0
	\end{equation}
    The validity region can be visualized in figure~\ref{fig:validity-region} for three different $Q/M$ values. Both $M$ and $r$ are given in geometrodynamic units in units of length. The dashed lines represent the $r_+$ horizon. We have also added the horizon for the $Q=0$ case as the solid black line for comparison. Note that for values of
    $r>2\;\lambda_{\textrm{compDE}}$, the Schwinger pair production rate~\eqref{eq:Schwinger} is always valid outside of the horizon. This can be understood by noting that equation~\eqref{eq:Schwinger_cond} can be rewritten as:
    \begin{equation}
		r > 2 \sqrt{1-\frac{2M}{r}+\frac{Q^2}{r^2}} \; \;\lambda_{\textrm{compDE}}.
	\end{equation}
    So certainly, if $r>2\lambda_{\textrm{compDE}}$, the Schwinger's calculation is applicable.  In order to obtain a bound on the black hole mass, let us go to the extremal case, which is the most constrained scenario of all.

	\paragraph{Exact result for the extremal case}
	$\;$\\
	If the black hole is exactly extremal, then equation~\eqref{eq:Schwinger_cond} can be rewritten as (note $r_+ = M$):
	\begin{equation}
		1-\frac{M}{r}  <  \frac{r}{2\lambda_{\textrm{compDE}}}\;, 
	\end{equation}
	which can be rewritten as:
	\begin{equation}
		r^2 - 2\lambda_{\textrm{compDE}}\;r+ 2M\lambda_{\textrm{compDE}} >0\;.
	\end{equation}
	Rearranging the terms we obtain:
	\begin{equation*}
		(r - \lambda_{\textrm{compDE}})^2 >\lambda_{\textrm{compDE}}\; (\lambda_{\textrm{compDE}}-2M)\;.
	\end{equation*}
	Note then that for $2M\geq \lambda_{\textrm{compDE}}$ this condition is automatically satisfied. This can be also visualized in figure~\ref{fig:validity-region}. In the dark sector, our simplified Schwinger condition then becomes:
    \begin{equation}
        M_\textrm{DE}  >  \frac{1}{2}\frac{m_{\textrm{Planck}}^2}{m_\chi} =  \frac{1}{2 \sigma_{m}}\frac{m_{\textrm{Planck}}^2}{m_e}\;.
    \end{equation}
    In terms of solar mass,
	\begin{equation}
		M_\textrm{DE}  \gtrsim \frac{1}{\sigma_{m}} 
  10^{-16} M_{\odot}\;.
	\end{equation}
	Note that for $\sigma_{m} \sim 10^{12}$, this condition allows us to use the Schwinger pair production rate for black holes as small as $10^{-28}M_{\odot} = 100$kg. Moreover, given that this condition is weaker than equation~\eqref{condition1}, the list of applicability conditions presented in section~\ref{S:updated-cond} is kept unaltered.

\subsection{Generalized evaporation time estimates}\label{S:near_extr_time}

Given the configuration space trajectory $Y(\mu)$, one can in principle estimate a black hole's total evaporation time by integrating equation~\eqref{eq:dmudtHWnew} for $\dif\mu/\dif\tau$. This, however, is rather nontrivial and generally requires a numerical integration. Another possible approach is to estimate the amount of time spent in each segment (mass or charge dissipation zones, near extremality and attractor region) separately. The integrals in the mass and charge dissipation zones in general still require a numerical integration. 
Fortunately, the scenario improves for both the approximate attractor  curve as well for near extremal evolutions.
This can be seen in figure~\ref{fig:tauattranalytic} in section~\ref{sec:timeest}, where we have performed the explicit integral for the approximate attractor curve and in equation~\eqref{eq:dTgen} in section~\ref{S:extremal}, where a straightforward analytical time estimate was obtained for the near extremal curve $Y=1$. 
Let us now extend these results to the scenario of dark electromagnetism.

\paragraph{Special case $(\mu_{\normalfont\textrm{h}}\ll z_0)$:} As discussed in section~\ref{S:small-y}, in this limit the Schwarzschild result is recovered. Therefore, the dark electromagnetism extension of equation~\eqref{eq:schwarzschild_lim} is simply given by:
\begin{equation}
    \Delta t = \frac{M_\textrm{s}^\chi}{s_0^\chi}\;\Delta\tau_\textrm{schw} 
    =  \frac{1920 \pi}{\alpha  \hbar }\left[\frac{32}{81}(M_\textrm{s}^\chi \mu)^3\right] = \frac{1920 \pi}{\alpha  \hbar }\left[\frac{32}{81}M^3\right] = \Delta t_\textrm{schw}\;,
    \label{eq:schwz_DE}
\end{equation}
where we have used the definition of $s_0$ given in equation~\eqref{eq:ConstDefs}. Note that one would normally expect the Schwarzschild evaporation time to depend on the electron mass. However, as discussed in footnote\footref{footnote}, in this work (as in the original HW article) we assume all mass lost in the Hawking process is due to the emission of massless particles --- implying an evaporation time independent of $m_\chi$ and $e_\chi$ whenever this limit is achieved~\footnote{This restriction to emission of massless particles would only have to be modified for extremely light black holes (\emph{i.e.,} very late in their evaporation), which are outside of the scope and goals of this study.}.

\paragraph{Attractor timescale}
In order to extend the obtained results for the scenario of dark electromagnetism, recall that equation~\eqref{eq:dtaumuattr1} for the (rescaled) time $\Delta\tau_\mathrm{att}(\mu = z_0)$ spent in the vicinity of the attractor curve depends solely on $z_0$. As shown in figure~\ref{fig:DE_attractors}, changing the properties of the dark electron implies a shift in the position  of $z_0$. On the other hand, in equation~\eqref{eq:z0fix} we have constructed specific rescaling formulae which keep the position of the attractor fixed, meaning $z^\chi_0=z_0$. This allows us to vary the dark-electron properties without having to worry about the position of $z_0$ in the new parameter space. Particularly, by using these rescaling relations it is sufficient to compute the time $\Delta\tau_\mathrm{att}(z_0)$ only once (keeping in mind that the corresponding physical time may then be obtained through a rescaling factor). The same is valid when calculating $\Delta\tau_\mathrm{att}(\mu)$ for any value of $\mu<z_0$.

\paragraph{Near-extremal solution}
For the near extremal time evolution, all we have to do is to replace the values of the electron charge and mass by their dark counterparts. In this case, given that $b_0^\chi$ is not kept fixed by the aforementioned rescaling equations, one may chose whatever rescaling is more suitable for any given problem. In this case, the dark electromagnetism extension for the near-extremal (rescaled) evaporation time is given by:
\begin{equation}
\Delta \tau^\chi = -\left. \frac{\exp[(\mu-z_0^\chi)\;b_0^\chi]}{b_0^\chi}\right|^{\mu_\textrm{f}}_{\mu_\textrm{i}} \;.
\label{eq:nearextrDE}
\end{equation}

\paragraph{Full evaporation starting from near-extremality}
As previously explained, $z_0$ marks an approximate location for the threshold between the attractor curve and the near-extremal phase (see section~\ref{S:approx-attractor}). Therefore, in order to calculate the full evaporation time starting from near-extremality, we must add the evaporation times in the near-extremal regime (from $\mu_\textrm{h}$ to $z_0$) and the time spent down the attractor curve (from $z_0$ to zero). The evaporation time estimate is then given by:
\begin{align}
\Delta \tau^\chi &= -\left. \frac{\exp[(\mu-z_0^\chi)\;b_0^\chi]}{b_0^\chi}\right|^{\mu_\textrm{f}}_{\mu_\textrm{i}} +\Delta \tau^\chi_{\mathrm{att}}(z_0),\\
&= \hphantom{-}\frac{1}{b_0^\chi} \left( \exp[(\mu_{\textrm{h}}-z_0^\chi) \;b_0^\chi] - 1\right) +\Delta \tau^\chi_{\mathrm{att}}(z_0).\label{eq:dTgenchi}
\end{align}
The first term corresponds to the time spent near $Y=1$, so we set $\mu_\textrm{i}=\mu_\textrm{h}$, with $\mu_\textrm{h}$ being the \enquote{hanging mass} defined in section~\ref{S:config-space-sol}, and $\mu_\textrm{f}=z_0^\chi$. In this case, whatever rescaling condition chosen, it must be the same for both regimes.

\paragraph{Special case $(\mu_{\normalfont\textrm{h}} - z_0^\chi\gg1/b_0)$:}
As in section~\ref{sec:timeest},  one may further simplify equation~\eqref{eq:nearextrDE} for the case where $\mu_{\textrm{h}} - z_0^\chi\gg1/b_0$ to: 
    \begin{equation}
		\Delta t^\chi 
        \sim \frac{M_\textrm{s}^\chi}{s_0^\chi \;b_0^\chi} \exp[(\mu_{\textrm{h}}-z_0^\chi) \;b_0^\chi]. \label{eq:dTgenXseg}
	\end{equation}
This simplification comes from the fact that when the system achieves near-extremality, its evolution slows down drastically, so that it spends the majority of its total evaporation time $t_\textrm{evap}$ near the $Y=1$ limit.
In this case, equation~\eqref{eq:dTgenXseg} provides an order-of-magnitude estimate for the total $t_\textrm{evap}$. A numerical verification of the preceding claim is presented in table~\ref{tab:DEresDiff}, where we show a comparison between analytical estimates and numerical results for different dark electron charge-to-mass ratios.

        \begin{figure}[ht]
		\begin{subfigure}{0.48\textwidth}
            \includegraphics[width=\linewidth]{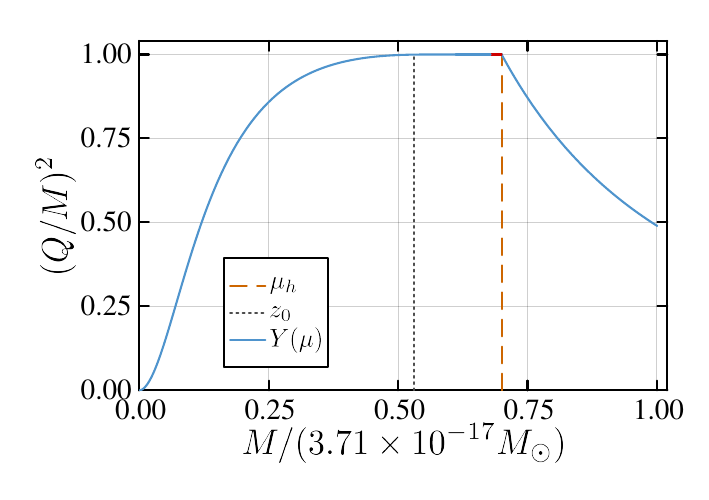}
            \end{subfigure}
            \begin{subfigure}{0.48\textwidth}
            \includegraphics[width=\linewidth]{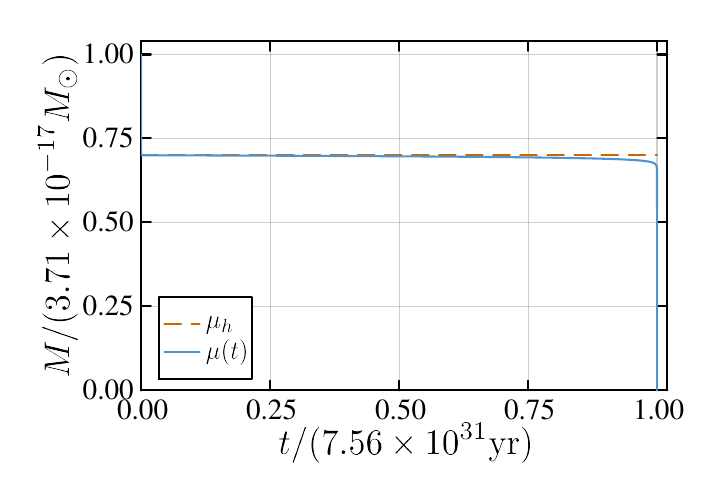}
            \end{subfigure}

            \smallskip
            \begin{subfigure}{0.48\textwidth}
            \includegraphics[width=\linewidth]{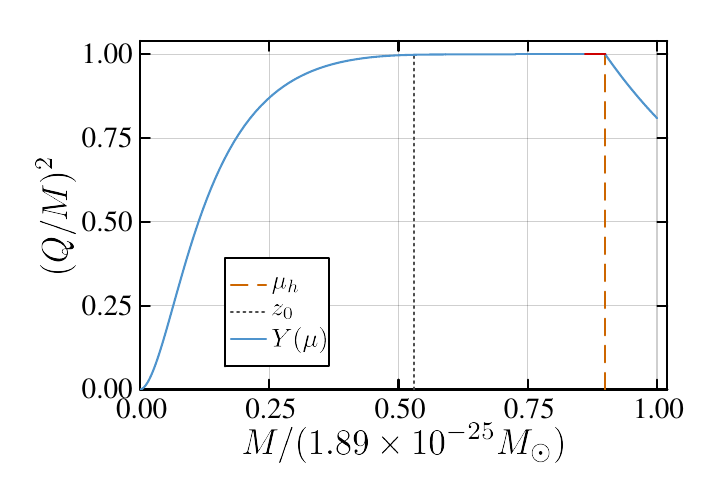}
            \end{subfigure}
            \begin{subfigure}{0.48\textwidth}
            \includegraphics[width=\linewidth]{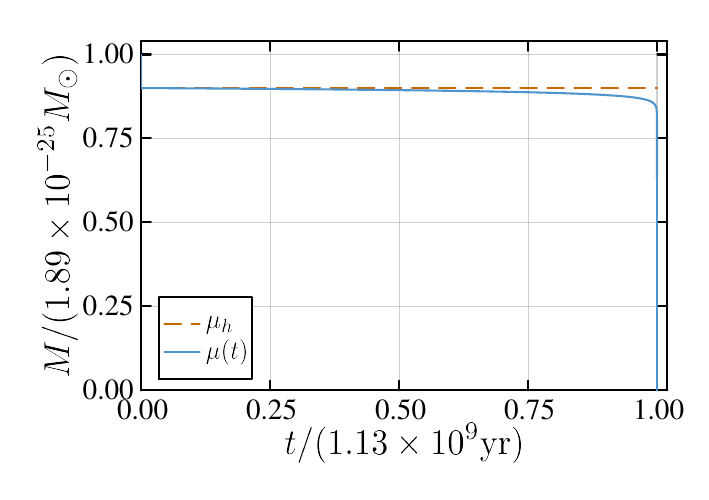}
            \end{subfigure}

        \caption{Configuration space (left column) and mass evolution (right column) plots for a solution with dark electron charge-to-mass ratio rescalings
        and hanging masses $(\xi, \mu_\textrm{h})$ given by $(10^{-14}, 0.7)$ (top row) and $(10^{-18}, 0.9)$ (bottom row). 
        In the configuration space plots, the vertical gray dotted line corresponds to $\mu=z_0$ and the segment indicated in red represents the region where the system spends $98\%$ of its evaporation time.}
        \label{fig:3runs}
    \end{figure}

  \subsubsection{\texorpdfstring{Numerical validation for $\mu_{\normalfont\textrm{h}}-z_0\gg1/b_0$}{Numerical validation for μh-z₀≫b₀}}
  \label{S:results}
\begin{table}[ht]
\begin{center}
\begin{tabular}{|c|c|c|c|c|c|}\hline
    $\xi$    & $\mu_{\textrm{h}}=0.70$ & $\mu_{\textrm{h}}=0.75$ & $\mu_{\textrm{h}}=0.80$ &  $\mu_{\textrm{h}}=0.85$ & $\mu_{\textrm{h}}=0.90$ \\ \hline
    $10^{-14}$   & $7.56\times 10^{31}$ & $4.01\times 10^{36}$ & $2.13\times 10^{41}$ & $1.13\times 10^{46}$ & $6.02\times 10^{50}$ \\ \hline
    $10^{-15}$   & $6.45\times 10^{23}$ & $9.08\times 10^{27}$ & $1.28\times 10^{32}$ & $1.80\times 10^{36}$ & $2.54\times 10^{40}$ \\ \hline
    $10^{-16}$   & $5.23\times 10^{15}$ & $1.94\times 10^{19}$ & $7.21\times 10^{22}$ & $2.67\times 10^{26}$ & $9.99\times 10^{29}$ \\ \hline
    $10^{-17}$   & $3.99\times 10^{7}$  & $3.84\times 10^{10}$ & $3.75\times 10^{13}$ & $3.67\times 10^{16}$ & $3.59\times 10^{19}$ \\ \hline
    $10^{-18}$   & $2.84\times 10^{-1}$ & $6.85\times 10^{1}$  & $1.73\times 10^{4}$  & $4.41\times 10^{6}$  & $1.13\times 10^{9} $ \\ \hline
\end{tabular}
\end{center}
\caption{Evaporation times (in years) as a function of the \enquote{hanging mass} $\mu_{\textrm{h}}$ and charge-to-mass ratio, where $\xi$ is the rescaling factor for the charge-to-mass ratio. In each case, the electron charge is rescaled by a factor $\sigma_e=10^{-4}$.} \label{tab:DEres}
\end{table}

\begin{table}[ht] 
\begin{center}
\begin{tabular}{|c|c|c|c|c|c|}\hline
    $\xi$     & $\mu_{\textrm{h}}=0.70$ & $\mu_{\textrm{h}}=0.75$ & $\mu_{\textrm{h}}=0.80$ &  $\mu_{\textrm{h}}=0.85$ & $\mu_{\textrm{h}}=0.90$ \\ \hline
    $10^{-14}$   & $2.29\times 10^{-5}$ & $2.36\times 10^{-5}$ & $2.10\times 10^{-5}$ & $1.89\times 10^{-5}$ & $1.72\times 10^{-5}$ \\ \hline
    $10^{-15}$   & $1.36\times 10^{-5}$ & $3.02\times 10^{-5}$ & $2.71\times 10^{-5}$ & $2.40\times 10^{-5}$ & $2.15\times 10^{-5}$ \\ \hline
    $10^{-16}$   & $7.93\times 10^{-5}$ & $3.72\times 10^{-5}$ & $3.76\times 10^{-5}$ & $3.29\times 10^{-5}$ & $2.90\times 10^{-5}$ \\ \hline
    $10^{-17}$   & $9.39\times 10^{-4}$ & $3.53\times 10^{-6}$ & $5.67\times 10^{-5}$ & $5.19\times 10^{-5}$ & $4.45\times 10^{-5}$ \\ \hline
    $10^{-18}$   & $5.16\times 10^{-2}$ & $2.36\times 10^{-3}$  & $1.09\times 10^{-5}$  & $1.07\times 10^{-4}$  & $9.32\times 10^{-5} $ \\ \hline
\end{tabular}
\end{center}
\caption{Relative difference in order-of-magnitude evaporation times between numerical results in table~\ref{tab:DEres} and the corresponding  estimates provided by the analytical formula~\eqref{eq:dTgenXseg}, according to Eq~\eqref{eq:tomdiff}.} \label{tab:DEresDiff}
\end{table}

    We now supply numerical validation for the claim that equation~\eqref{eq:dTgenXseg} provides a reliable order-of-magnitude estimate for the evaporation time. For this, we performed a total of $25$ runs, for rescalings of the charge-to-mass ratio  $\xi = \sigma_e/\sigma_m$ ranging from $10^{-14}$ to $10^{-18}$ (with charge rescaling parameter $\sigma_e=10^{-4}$), and hanging masses $\mu_{\textrm{h}} \in (\num{0.70},\num{0.90})$, respectively. Figure~\ref{fig:3runs} shows the results for two of these runs, displaying both the configuration space plots as well as the mass evolution for each case. In the configuration space plots, note the highlighted part of the curve in red, representing the region where the system spends $98\%$ of its total evaporation time. In the mass evolution plot, one can see a wide plateau, representing the concept of the hanging mass, in which the mass is roughly unchanged until the final stages of evaporation. The summarized results for the total evaporation times for the 25 runs are presented in table~\ref{tab:DEres}. Note that depending on the choices made for $\mu_{\textrm{h}}$ and $\varrho^\chi=\xi e/m_e$, the evaporation times vary by $\sim 52$ orders of magnitude. 
    
\begin{figure}[ht]
		\begin{subfigure}{0.48\textwidth}
            \includegraphics[width=\linewidth]{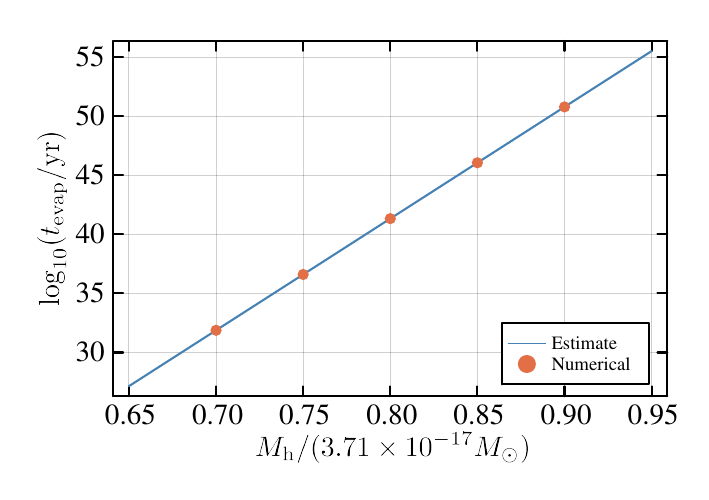}
            \end{subfigure}
            \begin{subfigure}{0.48\textwidth}
            \includegraphics[width=\linewidth]{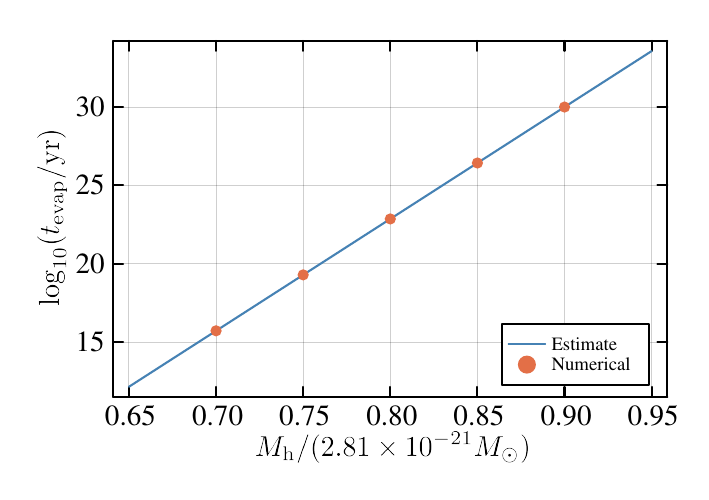}
            \end{subfigure}
            \caption{Evaporation time \emph{vs.} $M_\textrm{h}\defi \mu_{\textrm{h}} M_\textrm{s}$ (the \enquote{hanging mass}) for a charge-to-mass ratios $\xi=10^{-14}$ (left) and $\xi=10^{-16}$ (right). Numerical and estimated results are displayed; the estimate is given by equation~\eqref{eq:dTgenXseg}.} \label{fig:DETvM}
    \end{figure}

    The relative order-of-magnitude differences between the numerical results and equation~\eqref{eq:dTgenXseg} are given in table~\ref{tab:DEresDiff} and computed according to the formula:
        \begin{equation}\label{eq:tomdiff}
            {\Delta (\textrm{magnitude}) \defi }\frac{|\log_{10}(t_\textrm{analytical})-\log_{10}(t_\textrm{numerical})|}{|\log_{10}(t_\textrm{numerical})|}.
        \end{equation}
    The relative differences given in table~\ref{tab:DEresDiff} are minute, and demonstrate that for the region of parameter space considered, equation~\eqref{eq:dTgenXseg} provides a good estimate for the order of magnitude evaporation time. In figure~\ref{fig:DETvM}, we illustrate this with a logscale plot of evaporation time \emph{vs.} hanging mass $\mu_{\textrm{h}}$, and the reader can see that the numerical results are visually indistinguishable from the estimates provided by equation~\eqref{eq:dTgenXseg}. Again, the numerical codes and scripts can be found on the Github repository \url{https://github.com/justincfeng/bhevapsolver}.

\enlargethispage{10pt}

\subsection{Mass bounds for dark-charge PBHs}
\label{S:Muniv}

In this section, our aim will be to extend the bounds imposed (due to Hawking radiation) on the allowed fraction of dark matter {in the form of} PBHs. As we have seen, the presence of even a small amount of (dark) charge is capable {of extending} the lifetime of black holes by many orders of magnitude. Using the results {derived} so far, we will now calculate the minimum mass {$M_\textrm{univ}$} for a PBH to live longer than the age of the universe {$t_\textrm{univ}$}. 

\enlargethispage{20pt}

\begin{figure}[!ht] 
	\centering
\includegraphics[width=\linewidth]{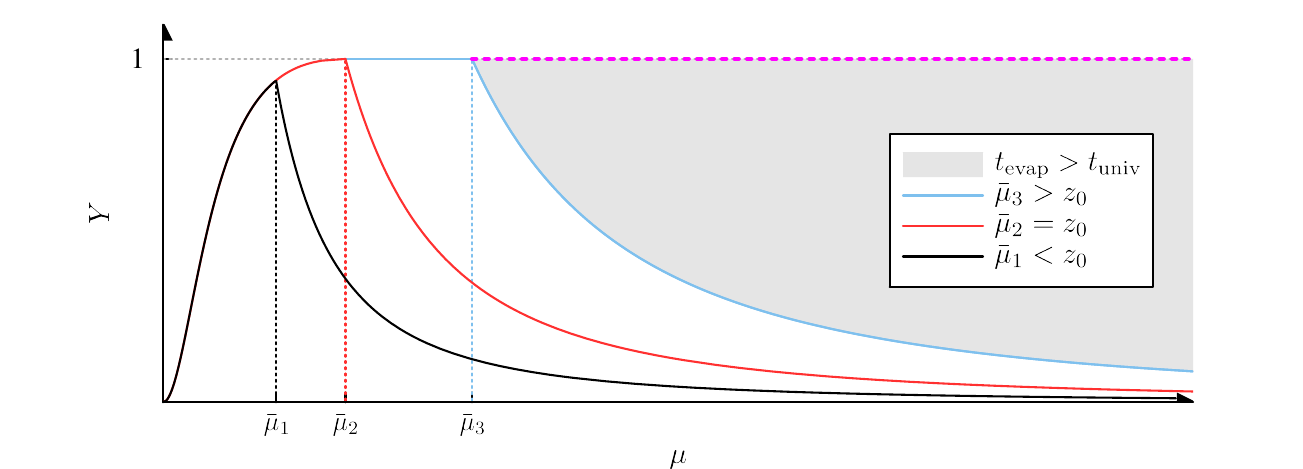}
    \caption{Illustration of three different solutions in configuration space, distinguished by values of rescaled mass $\bar\mu$ at which the system either reaches near-extremality or the attractor region. The region of the configuration space in which a near-extremal PBH spends a time longer than the age of the universe  $t_\textrm{univ}$ in near-extremality, is shaded in gray. 
    Evaporation times along the magenta line satisfy $\Delta t_\textrm{evap}>t_\textrm{univ}$ for PBHs with $\bar\mu - z_0 \gg 1/b_0$.}\label{fig:psmubar}
\end{figure}

In the previous section, we have presented {results} for the total evaporation time {of} a black hole evolving along the approximate attractor curve and {of} the near-extremal {cases}. Recall that the condition for a black hole to achieve near extremality is $\mu_{\textrm{h}}\geq z_0$, where $\mu_{\textrm{h}}$ is the {previously defined} hanging mass. In the {following} analysis, we will consider the evolution time only along {these} two phases (attractor and near extremality), without taking into account the evaporation time {spent in} the mass dissipation zone --- as we will show, the {relevant} region where {this has to be taken} into account is {small} in parameter space. Nevertheless, this means that all mass bounds presented in this section are mildly, if not highly conservative, depending on whether near-extremality is achieved or not. A more in-depth discussion of this follows below. For now, let us divide the evaporation regimes into four different cases (see equations~\eqref{eq:schwz_DE}--\eqref{eq:dTgenXseg}): 
\begin{equation}
\Delta t_\textrm{evap}(\bar\mu) = \frac{M_s^\chi }{s_0^\chi b_0^\chi }\cdot
\begin{cases} 
    \Delta \tau_\textrm{schw}(\bar\mu) \;, & \bar\mu\ll z_0^\chi\\
      \Delta \tau_{\textrm{att}}(\bar\mu) \;, & \bar\mu\leq z_0^\chi\\
       \left( \exp[(\bar\mu-z_0^\chi) \;b_0^\chi] - 1\right) + \Delta \tau_{\textrm{att}}(z_0) \;,& \bar\mu> z_0^\chi\\
      \exp[(\bar\mu-z_0^\chi )\;b_0^\chi ]\;,& \bar\mu - z_0^\chi \gg 1/b_0^\chi
   \end{cases}\;
   \label{eq:evap_time_cases}
\end{equation}
where $\bar\mu$ is defined as the rescaled mass where the system either becomes near-extremal (in which case $\bar\mu=\mu_{\textrm{h}}$) or reaches the attractor curve. Figure~\ref{fig:psmubar} illustrates the system's evolution and the location of $\bar\mu$ for three different cases.
Now, setting $\Delta t_{\textrm{evap}} = t_\textrm{univ} = \num{13.8e9}$ years and solving for $\bar\mu$, we obtain the following solutions for the minimum mass $M_\textrm{univ} = M_\textrm{s}\;\bar\mu$:
\begin{equation}
M_\textrm{univ} = 
\begin{cases} 
    M_\textrm{schw}\;, & \bar\mu\ll z_0^\chi\\
      M_\textrm{att}^\textrm{univ}, & \bar\mu\leq z_0^\chi \\
    M_\textrm{near-extr}^\textrm{univ} \;,& \bar\mu> z_0^\chi \qquad\\
     M_\textrm{extr}\;,& \bar\mu - z_0^\chi \gg 1/b_0^\chi
   \end{cases}
   \label{eq:Munivmin}
\end{equation}
Let us now {address} this case by case.

\paragraph{Special case $(\bar\mu - z_0 \gg 1/b_0)$:} For the special case when the $\bar\mu = \mu_{\normalfont\textrm{h}}$ is significantly larger than $z_0$, the evaporation time along the attractor region will be many orders of magnitude smaller than the near-extremal evolution. The same happens in the mass dissipation zone. Numerical results demonstrating this claim may be be found in section~\ref{S:results}, where we show that in such cases the black hole spends \SI{98}{\percent} of its total evaporation time in the vicinity of the hanging mass. Hence, one may consider the following approximation for the total evaporation time:
\begin{equation}
    \Delta t_\textrm{evap}(\bar\mu) = \frac{M_s}{s_0b_0}\;\exp[(\bar\mu-z_0)\;b_0]  \;,\quad\qquad\bar\mu\gg z_0 \;.
    \label{eq:muh_gg_z0}
\end{equation}
Equation~\eqref{eq:muh_gg_z0} allows for a simple analytical solution. Setting $\Delta t_\textrm{evap}=t_\textrm{univ}$, we obtain a lower bound for the mass of near-extremal PBHs which would not yet have fully evaporated by the present day: 
\begin{equation}\label{eq:Muniv}
    M^\chi_\textrm{extr} \defi \frac{e_{\chi}}{m_\chi^2 }\frac{\hbar}{\pi} 
    \ln\left[\frac{e_\chi^3 t_\textrm{univ}}{2 \pi ^2 \hbar ^2}\right]=
    \frac{e \hbar c  \ln \left(\frac{e^3 c t_\textrm{univ}}{16 \pi ^{7/2} \hbar ^2 \sqrt{\varepsilon_0^3 G}}\right)}{2 \pi ^{3/2} m_e^2 \sqrt{\varepsilon_0 G^3}},
\end{equation}
where in the last equality we have recovered SI units for completeness. Also, aiming to keep a light notation, and given that from now on we will always be dealing with dark electromagnetism, we will just refer to $z_0^{\chi}$ as $z_0$ and $M^\chi_\textrm{extr}$ as $M_\textrm{extr}$.

In order to understand equation~\eqref{eq:Muniv}, in figure~\ref{fig:psmubar} we have depicted $\mu_{\textrm{extr}} = M_{\textrm{extr}}/M_\textrm{s}$ by $\bar\mu_3$ corresponding to the fourth case of~\eqref{eq:evap_time_cases}.  
Any PBH which is formed with an initial mass and charge such that it lies in the gray region of the configuration space, will take longer than the age of the universe to evaporate. This comes from the fact that the evaporation times for line segments starting at any point on the magenta dashed line and finishing at $z_0$ satisfy $t_\textrm{evap}>t_\textrm{univ}$.

Given that equation~\eqref{eq:Muniv} for $M_\textrm{extr}$ is applicable only for $\bar\mu - z_0 \gg 1/b_0$, $M_\textrm{extr}$ is the minimum initial mass a \emph{near-extremal} black hole needs to have in order to spend a time equal to the age of the universe in near-extremality. 
For the case of standard electromagnetism, this mass is approximately $M_\textrm{extr} \approx \num{2.2e7} M_\odot$~\footnote{Note that this value is much greater than the usual Schwarzschild limit. This, however, does not come as a surprise, especially since, as mentioned in text, the definition of $M_{\textrm{extr}}$ does not take the evaporation time past $z_0$ into account. 
}. Looking at figure~\ref{fig:configuration-space}, one can see that this mass is actually much below the region where near-extremality is achieved (meaning $\mu_\textrm{extr} < z_0$). 
This means that a RN black hole with a \emph{standard} --- not dark ---, electromagnetic charge and a mass $M_{\textrm{extr}}$ would actually \emph{never reach near-extremality} via the Hawking and Schwinger evaporation alone. This implies that, for standard model electromagnetism, neither equation~\eqref{eq:Muniv} nor equation~\eqref{eq:muh_gg_z0} are applicable. We will return to this point later in more detail, but for now reader may find the region of the parameter space where this regime starts (and dominates) in the lower right corner of figure~\ref{fig:DE_mass_charge}, to the right of the solid line marking $M_\textrm{univ}\approx M_{z_0}$.

\paragraph{Near-extremal case $(\mu_{\normalfont\textrm{h}}>z_0)$:} 
Recalling the discussion in section~\ref{S:approx-attractor} and remembering that $M_{z_0}\defi z_0 M_\textrm{s}$, black holes which do become near-extremal must satisfy $M_\textrm{s} \mu_\textrm{h} \geq M_{z_0}$. In this way, the domain of validity  for equation~\eqref{eq:Muniv} can be restated as $M_\textrm{extr} \gg M_{z_0}$. When this is not the case, one must take the evaporation time along the attractor region into account:
\begin{equation}
    \Delta t_\textrm{evap}(\mu) =  \frac{M_s}{s_0b_0}\;\exp[(\mu-z_0)\;b_0] + \Delta t_{\textrm{att}}(z_0)\;.
\end{equation}
By solving this equation for $\mu$ with $\Delta t_\textrm{evap} = t_\textrm{univ}$, one can find the minimum mass for the near-extremal case. As an analytical expression for this case 
would be too complicated to present here, let us simply label this solution by $M_\textrm{near-ext}^\textrm{univ}$ for clarity in future discussions.

 \begin{figure}[t]
	    \centering
		\includegraphics[width=0.65\linewidth]{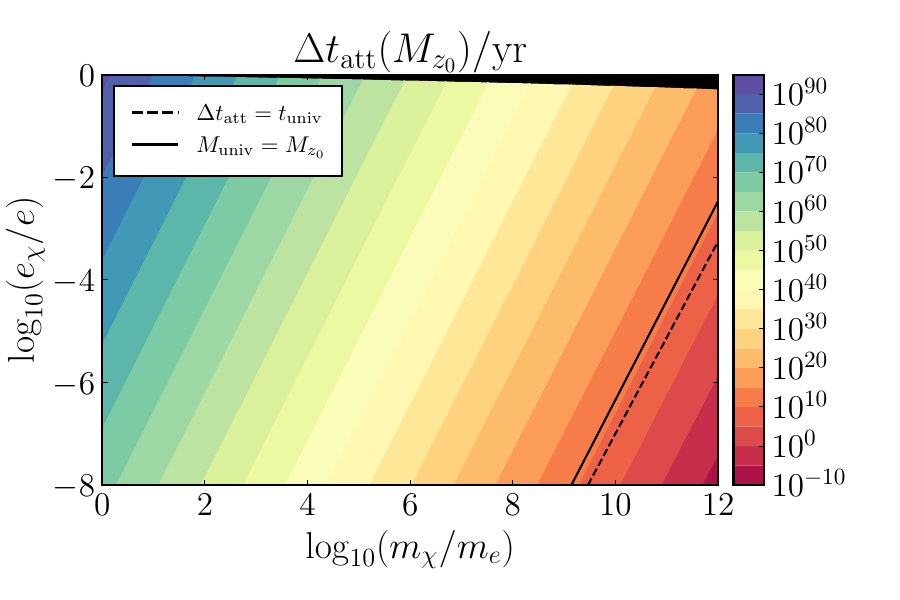}
		\caption{Contour plot of the evaporation time estimate $\Delta t_{\mathrm{att}}$ along the attractor from $\mu \sim z_0$ to $\mu=0$.
        The dashed black line represents the location where $\Delta t_{\mathrm{att}} = t_{\mathrm{univ}}$ and the solid black line represents the location where $M_\textrm{univ} = M_{z_0}$.
        The lowest value for $\Delta t_{\mathrm{att}}$ occurs on the lower right corner of the plot. Although not visible in this plot, we remind the reader that the mass $M_{z_0}:=M_\textrm{s}z_0$ depends on $e_\chi$ and $m_\chi$ (see equation~\eqref{Eq:z0chi}) and its lowest value is also at the right bottom corner of the plot. The black area near the top is the region of parameter space excluded by condition~(ii) (equation~\eqref{eq:condition2}).
        } \label{fig:DE_mass_charge_time}
    \end{figure}

\paragraph{Full attractor evolution $(\mu_{\normalfont\textrm{h}} = z_0)$:} 
We now aim to understand the orders of magnitude of the evaporation timescales and how these are affected by varying the dark electron's properties. For this purpose, we have computed for different scenarios of dark electromagnetism the time for a black hole to evaporate, starting at the beginning of the attractor ($\mu_{\normalfont\textrm{h}} = z_0$). 
In figure~\ref{fig:psmubar}, this would correspond to the curve designated by $\bar\mu_2=z_0$.
The solution for the rescaled time $\Delta\tau_\textrm{att}$ is presented in equation~\eqref{eq:dtau(mu)}. Keeping in mind that the physical time is given by $\Delta t_{\mathrm{att}}(z_0) = \Delta \tau_{\mathrm{att}}\;M_\textrm{s}/s_0^\chi$, we have made use of the rescaling relations~\eqref{eq:z0fix} in order to extend the electromagnetic solution to different scenarios of dark electromagnetism (note that, using those relations, $z_0$ is kept fixed). The result is presented in figure~\ref{fig:DE_mass_charge_time}. The dashed black line represents the location where $\Delta t_{\mathrm{att}} = t_{\mathrm{univ}}$, meaning that values of $(m_\chi, e_\chi, M_{z_0})$ to the left of this line  take longer than $t_{\mathrm{univ}}$ to evaporate. The solid black line represents the location where $M_\textrm{extr} = M_{z_0}$, with values to the left of this line having $M_\textrm{extr}<M_{z_0}$. The fact that this line is placed to the left of the $\Delta t_{\mathrm{att}} = t_{\mathrm{univ}}$ threshold guarantees that whenever $M_\textrm{extr}<M_{z_0}$, one may safely take $M_{z_0}$ as the new lower limit for the PBH mass $M_{\textrm{univ}}$.

\begin{figure}[t]
	    \centering
		\includegraphics[width=0.65\linewidth]{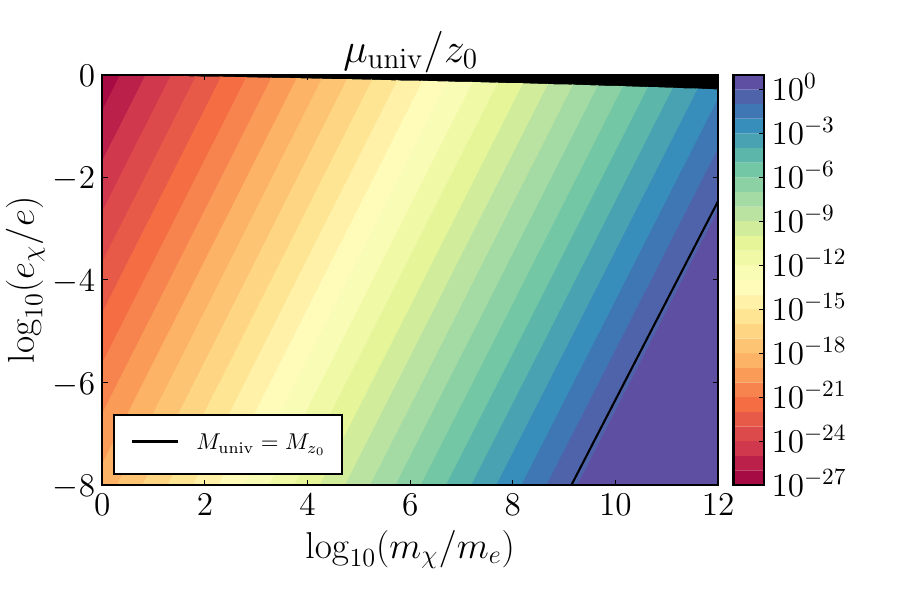}
		\caption{Contour plot of the fraction $\mu_\textrm{univ}/z_0$. This illustrates how the region where taking the mass dissipation zone time into account $(10^{-1}\lesssim \mu_\textrm{univ}/ z_0 \lesssim 1)$ is in a sufficiently constrained region of parameter space. As in figure~\ref{fig:DE_mass_charge_time}, the black area near the top is the region of parameter space excluded by condition~(ii) (equation~\eqref{eq:condition2}). The lowest values of $\mu_\textrm{univ}/z_0$ occur in the top left.} \label{fig:mu_over_z0}
\end{figure}

\begin{figure}[ht]
	    \centering
        \includegraphics[width=0.65\linewidth]{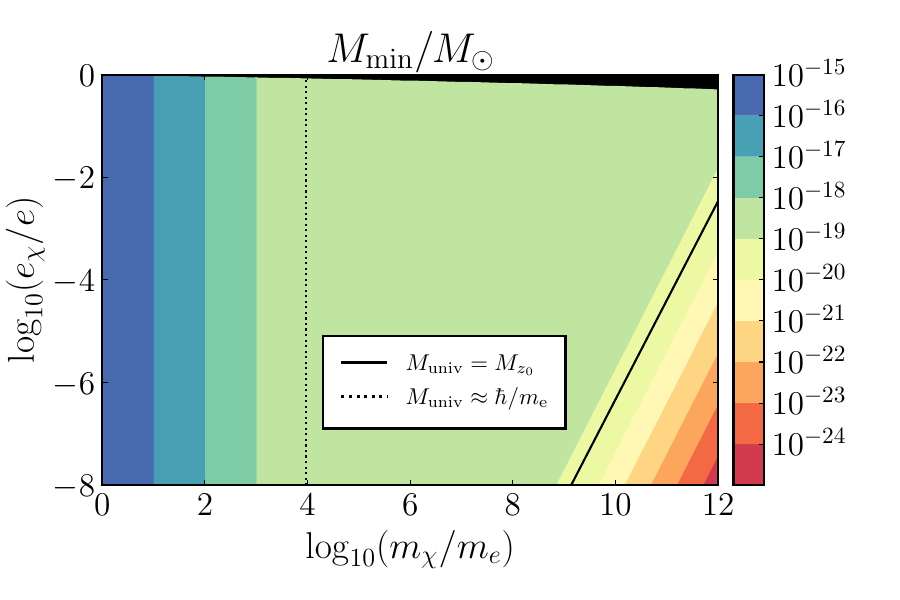}
		\caption{Contour plot showing $M_\textrm{univ}$ for different scenarios of dark electromagnetism. The vertical dotted line represents the minimum mass allowed by condition~(i) (equation~\eqref{condition1}). The solid line to the right is the location where $M_{\mathrm{univ}} = M_{z_0}$. As in figures~\ref{fig:DE_mass_charge_time} and~\ref{fig:mu_over_z0}, the black area near the top is the region of parameter space excluded by condition~(ii) (equation~\eqref{eq:condition2}). The lowest values of $M_\textrm{univ}$ occur at the left.
        }
        \label{fig:DE_mass_charge}
\end{figure}

\paragraph{Attractor evolution $(\mu_{\normalfont\textrm{h}}<z_0):$}
Now we will look at the more general case when a black hole, evolving from the mass dissipation zone, does not achieve near extremality, but instead \enquote{hits} the attractor curve at a determined, subextremal mass $\mu_{\textrm{att}}<z_0$.
In figure~\ref{fig:psmubar}, an example of this case is represented by the curve designated by $\bar\mu_1$.
In this case, one may calculate the evaporation time along the attractor region as described in sections~\ref{sec:timeest} and~\ref{S:near_extr_time}. 
Given that we have chosen the auxiliary functions for curve fitting $\Delta\tau_\textrm{att}(\mu)$~\eqref{eq:dtau(mu)} to be easily invertible, this allows us to solve them for the value of
$\mu$ which satisfies $\Delta t_{\textrm{att}} = t_\textrm{univ}$ (keeping in mind the scaling transformations between $t$ and $\tau$):
\begin{equation}
    \mu_\textrm{att}^\textrm{univ} = \left(  \frac{b s_0^\chi t_\textrm{univ}}{M_\textrm{s} -as_0^\chi t_\textrm{univ}}\right)^{1/m}\;.\label{eq:Muniv-non-extremal-mh-less-z0}
\end{equation}
Here, $(a,b,m)$ are the best-fit parameters of section~\ref{sec:timeest}.
Multiplying this equation by the mass scale, we have $M_\textrm{att}^\textrm{univ} = M_\textrm{s} \;\mu_\textrm{att}^\textrm{univ}$. Since the time along the mass dissipation zones is not taken into account when deriving this formula, this is a conservative lower bound for the mass $M_\textrm{univ}$ of non-extremal black holes.

\paragraph{Limiting case $(\mu\ll z_0)$:}
When calculating the minimum mass $M_\textrm{univ}$ that a PBH must have in order to live longer than the age of the universe, a surprisingly large region of the parameter space $(m_\chi, e_\chi, M_\textrm{PBH})$ is actually covered by the limiting case $(\mu\ll z_0)$. This can be seen in figure~\ref{fig:mu_over_z0}, where we have plotted the fraction $\mu_\textrm{univ}/z_0$ for different values of dark charge and mass. In terms of the dark electron mass, note that only for values of $m_\chi>10^8 \;m_e$ should one start worrying about departing from the low mass limit. Linking this to the time estimation for small $\mu$ presented in section~\ref{S:near_extr_time}, we see that the Schwarzschild limit is actually a good approximation for the minimum mass in a vast region of the parameter space here analyzed.

\paragraph{General case:} Having now looked more carefully at each individual case identified in equation~\eqref{eq:Munivmin}, let us further collect these results in order to define a conservative lower bound valid across all cases considered for the mass $M_{\mathrm{univ}}$ of those dark-charged PBHs which would not yet have fully evaporated by the present day. In figure~\ref{fig:DE_mass_charge} we have created a contour-plot of $(m_\chi, e_\chi, M_{\textrm{min}})$. Here, $M_{\textrm{min}}$ is defined as $ M_{\textrm{univ}}$, except for the regions where condition~(i) and condition~(ii), given by equations~\eqref{condition1} and~\eqref{eq:condition2}, respectively, are no longer valid. In this case, we have plotted the minimum mass allowed by condition~(i) (region to the left of the vertical dashed line) and simply excluded the region where condition~(ii) is violated. However, as we have just argued, the true value in these regions simply reduces to that of a Schwarzschild black hole. This is also the reason for the large plateau region in the center of the figure. Note that once one starts to move away (to the left) from the $M_{\textrm{univ}} = M_{z_0}$ line, the Schwarzschild result is quickly achieved and maintained (compare also with figure~\ref{fig:mu_over_z0}). On the other hand, once one moves to the right of this line, the near-extremal regime starts to take over, quickly dropping the minimum mass bound. 
Figure~\ref{fig:DE_mass_charge} allows us to understand how much one must vary $m_\chi$ and $e_\chi$ in order to have low mass PBHs which have not yet completely evaporated.

\subsection{Model building considerations}

It is important to note that, the seemingly large ranges in mass $m_\chi$ or charge $e_\chi$ should not be discouraging. Already in the standard model of particle physics (mildly modified to accommodate neutrino masses), we have enormous differences in mass scales: Neutrino masses are at most of the order \SI{e0}{\electronvolt}, while the tauon has a mass of \SI{1.77686}{\giga\electronvolt} (so $m_\tau/m_\nu \sim 10^{12}$). Hence, the mass ranges seen in our model would not be outlandish from the perspective of those seen in the standard model for elementary particles. Meanwhile, discounting uncharged particles for the moment, $\log_{10}(q/e)$ for particles of charge $q$ in the standard model is close to $0$ (\emph{i.e.,} $q/e$ is close to unity). However, the plateau of figure~\ref{fig:DE_mass_charge} extends to this value, and from the standard model itself it is much harder to develop intuition about \emph{different} $U(1)$ coupling strengths. Neither colour charges nor hypercharges are easily compared to electromagnetic charges, and thus are of little help for this purpose.

\section{Conclusions}\label{S:conclusions}

In this paper, we considered the evaporation of charged (Reissner--Nordström) black holes. We have reviewed and extended the original results obtained by Hiscock and Weems \cite{Hiscock:1990ex} to PBHs with a dark $U(1)$-charge, and evaluated how varying the dark electron mass $m_\chi$ and charge $e_\chi$ affect the PBH evolution. 

In their original work, HW have shown that Reissner--Nordström black holes do not always evolve as one naively might think --- a quick discharge to a Schwarzschild black hole. 
While in the standard electromagnetic scenario this is true for low mass black holes, (isolated) black holes above a certain mass may present interesting and unexpected evolution scenarios --- for example, a black hole with an initial charge-to-mass ratio as low as $1\%$ may naturally evolve to a near-extremal state, as represented in figure~\ref{fig:configuration-space}. 

The unexpected behavior arises from the interplay of two fundamentally distinct quantum processes governing the evaporation of Reissner–Nordström black holes: Hawking radiation and the Schwinger pair-production effect. The common existence of these two quantum effects leads to a $(M, (Q/M)^2)$ configuration space split into two regions --- the mass and the charge dissipation zones --- each one dominated by one of the quantum processes mentioned above (see figure~\ref{fig:configuration-space} and section~\ref{S:config-space}). These two regions are divided by an \enquote{attractor curve} or attractor region, whose location on the configuration space depends on the mass and charge of the lightest fermionic particle carrying a non-zero charge. In the standard electromagnetism case, the location of this attractor region is such that only black holes of masses greater than $\approx 5 \times10^7 M_\odot$ may naturally achieve near-extremality along its evolution. 

In our analysis, we have rewritten the original HW equations in terms of new variables, which serve to clarify the behavior of the system and facilitate its numerical implementation. 
These have also allowed us to obtain
for the first time a closed form analytical expression for the approximate attractor curve (see equation~\eqref{eq:muatt}), which very clearly highlights how the attractor curve depends on the dark $U(1)$'s parameters (via $z_0$).
We have also presented clear and simple expressions for approximate configuration space evolution $Y(\mu)$ in both the mass and charge dissipation zones. 
Furthermore, we have obtained approximate analytical expressions for the evaporation time estimate along the attractor region, for the life-time of near-extremal black-holes and in the low $(Q/M)$ limit. For the near-extremal case, a comparison between the approximate solution time and the full numerical solution can be found in table~\ref{tab:DEresDiff}.

We then extended these results to a scenario of dark electromagnetism, tentatively modelled on the Lagrangian~\eqref{eq:lagrangian} with a massless, uncharged \enquote{dark photon}, and a \enquote{dark electron $\chi$} of mass and charge $m_\chi$ and $e_\chi$, respectively. Even if such a $U(1)$ is not part of the current universe, PBHs formed much earlier could have retained such a \enquote{dark charge}. This allows them to be a promising PBH dark matter candidate, in counterpoint to their standard electromagnetic RN counterparts. 

To investigate this possibility, we have updated the validity conditions of the model adopted by the original HW analysis, as well as the Schwinger pair-production results for the case of dark electromagnetism. We then explored the dependence of the location of the attractor curve on the dark electron's properties, showing that by increasing the dark electron's mass $m_\chi$ and/or lowering its charge $e_\chi$, one can push the location of the attractor curve to significantly lower black hole masses $M$ (see figures~\ref{fig:DE_attractors}). We have also extended the evaporation time estimates obtained for the standard electromagnetism case to the dark sector. 

Existing results for constraints on the mass $M_{\textrm{PBH}}$ allowed for Schwarzschild black holes as candidates for PBHs contributing to dark matter were shown in figure~\ref{fig:PBH4}. 
Aiming to update this mass bound to our dark-charged PBH proposal, we have used the lifetime estimates obtained above, and explored the region of parameter space $(m_\chi, e_\chi, M_\textrm{PBH})$ in which dark-charged PBHs have a lifespan that is longer than the age of the universe. To do so, in equation~\eqref{eq:evap_time_cases}, we have defined four cases of (rescaled) black hole masses each of which could give the most stringent bound on the mass $M_{\textrm{univ}}$ (within the region of validity of the HW model) as per equation~\eqref{eq:Munivmin}. 
The results are summarized in figure~\ref{fig:DE_mass_charge}, demonstrating that, depending on the values of $e_\chi$ and $m_\chi$, the minimum mass $M_\textrm{univ}$ for PBHs which will not yet have fully evaporated by today can go \emph{at least} as low as $10^{-24} M_\odot$.
This is an important step for understanding the importance of PBHs as dark matter candidates. Modelling PBHs solely on the Schwarzschild solution \emph{cannot} give a comprehensive picture, and the resulting, extant constraints on the fraction of dark matter composed of PBHs are of limited use. 
At the very least the full extent of the Kerr--Newman family should be further explored, including the many physical effects that influence the evaporation process besides the Hawking effect --- as in the present article, the Schwinger effect.

\section{Future directions and open questions}

\paragraph{More general classes of black holes} 
Most immediately, this leads to the question of angular momentum. Even before adding charge to PBHs, rotating black holes (described in general relativity by the Kerr solution) are considered to be the \enquote{standard black hole}~\cite{KerrFest,Visser:2007fj,Adamo:2014baa,ONeill:2014rav}. Non-rotating black holes are not observed to date. As electromagnetic potentials enter the evaporation in a similar way to angular momentum, these will likewise already modify constraints derived from simple models of evaporating Schwarzschild black holes. However, there is no analogue of the Schwinger effect in this case, though superradiance enters the picture \cite{Frolov:1998wf,Brito:2015oca}. Adding angular momentum to the charged black holes considered in the present article would lead, within general relativity, to the Kerr--Newman family of black hole solutions. These are parametrized by charge, angular momentum and mass.\footnote{Though magnetic charge is also easily accommodated, and has at least some similarity to the considerations of the present article.} As our present discussion was wholly concerned with non-rotating black holes, already the question of whether our results obtained for RN black holes change noticeably when including rotation presents itself as a future extension of our dark electromagnetic model. At the very least, the HW model for the evaporation of charged, \emph{non-rotating} black holes would need extension to the full Kerr--Newman family, independent of whether or not \enquote{dark charge} is included.

Beyond rotation, there are few limits to one's imagination on how to open up parameter space for PBHs as dark matter candidates, when other concerns of particle physics need to entertain changes to the standard model (for example, those discussed in chapters~85 to~95 of \cite{ParticleDataGroup:2024cfk}): Changes to the standard model would also imply changes to the parameters encountered in the metrics describing black holes. Depending on the changes (as the above allusion to string theory or other higher-dimensional theories should make clear), different \emph{types} of black holes beyond the Kerr--Newman family would arise \cite{Delgado:2016jxq,Brito:2018hjh,Fernandes:2020gay,Gomez:2023wei}. Besides questions of astrophysics around these black holes, this also leads to many more options for these black holes to form a (more or less) significant fraction of the dark matter content of the universe. Finally, many modifications of the standard model undertaken in search of a theory of quantum gravity dabble in higher dimensions, and the HW model has already been extended to this case \cite{Xu:2019wak}. Adding extra dimensions can change things drastically (for \enquote{normal} and PBHs alike), as unlike in $3+1$ dimensions, more axes of rotation need to be specified for rotating black holes.

In future studies, one may consider the implications of a change in charge-to-mass ratio on other astrophysical processes potentially applicable in our scenario --- see, for example, \cite{Moreno:2016urq,Feng:2022evy}. Since we are concerned with the late stages of black hole evaporation, it is perhaps also worth comparing the observational implications of our scenario with that of black hole remnants (we refer the reader to the reviews \cite{Ong:2024dnr} and \cite{Chen:2014jwq}) or other exotic proposals for the end state of black hole evaporation \cite{Hayward:2005gi,Hossenfelder:2009xq,Bianchi:2018mml,Feng:2023klt}.

\paragraph{More general dark matter models} 
While more general classes of black holes influence the gravity side of our analysis, similarly, the particle physics side can be modified: For example, 
allowing for additional couplings to other particles or, for example, non-linear (dark) electromagnetism \cite{Sorokin:2021tge}. As previously mentioned, it is important to understand how having a more complete and realistic dark $U(1)$ model would impact the dynamics and evolution of dark-charged PBHs. This will also impact the question of overcharging, to be mentioned further below.
Moreover, a more complex but still feasible path is to go beyond the $U(1)$ charge, and extend the analysis to non-abelian charges. Spherically symmetric solutions of the Einstein--Yang--Mills equations have been considered in the literature \cite{Yasskin:1975ag,Herdeiro:2017oxy,Gomez:2023qyv}, and also the Schwinger effect for the Yang--Mills case \cite{Ragsdale:2017wgi}; it should be a relatively straightforward task to extend the Hiscock and Weems analysis to the Yang--Mills setting.

Another interesting avenue of research would be to extend the ideas of references \cite{Cheek:2021odj,Hooper:2019gtx,Cheek:2021cfe} to the model proposed here --- investigating and constraining the dark matter relic density in models in which a significant part of the dark matter currently in our universe is emitted by PBHs themselves. As we have seen throughout our analysis, the evaporation process is even sparser \cite{Gray:2015pma} than that of Schwarzschild black holes. With dark matter particles being emitted (primarily) very close to the PBH's horizon, such dark matter will be more difficult to detect and therefore to constrain at least directly, if not also indirectly.

\paragraph{Overcharging and cosmological censorship} 
If near-extremal dark-charge black holes are present in the universe, it is perhaps worth remarking on processes that can potentially overcharge the black hole by adding small amounts of angular momentum or charge \cite{Cardoso:2018nvb,Davey:2024xvd}. While it has been shown under some rather general assumptions that a Kerr--Newman black hole cannot be overcharged \cite{Wald:1974hkz,Sorce:2017dst}, those analyses only consider a single $U(1)$ gauge field. Near-extremal dark-charge black holes, on the other hand, can in principle absorb standard model electrons, which do not experience a strong electrostatic repulsion from dark $U(1)$ charges. In \cite{Izumi:2024rge}, multiple $U(1)$ gauge fields were considered in the test particle approximation, and while their analysis does not rule out the possibility of overcharging, the authors strongly caution that a general analysis (perhaps along the lines of \cite{Sorce:2017dst}) is needed to conclusively determine whether multiple $U(1)$ gauge fields can lead to a violation of the weak cosmic censorship conjecture.

\paragraph{Dark-charged PBH formation} 

Note that in the current article we have focussed on the evaporation of dark charge PBHs, side-stepping (for now) the question of their formation.
This is a fascinating question which we will not (yet) attempt to answer, but we will not refrain from pointing out several possible formation mechanisms. 
Overall PBH formation is a relatively well-studied topic, see for instance \cite{LISACosmologyWorkingGroup:2023njw}.

First of all, besides the most common approach for the formation of PBHs being the collapse of primordial density fluctuations, this is not the only path. Another interesting mechanism comes from the collapse of cosmic strings originating in the early universe \cite{Hawking:1987bn, Polnarev:1988dh}. As has been shown, these cosmic strings may have a charge \cite{Weigel:2010zk, Oliveira:2012nj}. Therefore, one possible scenario for the formation of dark-charged PBHs could come from the collapse of dark-charged $U(1)$ strings. Another possibility comes from the formation of PBHs during a cosmic phase transition, for example, the collapse of Fermi balls \cite{Kawana:2021tde, Huang:2022him}. Since these Fermi balls may also carry a charge, these could generate dark charged PBHs as a final product. This mechanism was already discussed in detail in \cite{Morris:1998tw}. One may also consider PBH formation from Q-balls \cite{Cotner:2016cvr}, which again may be charged \cite{Benci:2010cs}.

Focusing now on the formation of PBHs due to the collapse of density fluctuations, another fundamental difference comes into play. While one does not expect the formation of charged PBHs in the scenario of standard electromagnetism due to the electric repulsion forces between two electrons being many orders of magnitude stronger than their gravitational attraction, this also changes in the model here proposed. As discussed throughout the paper and presented in figure~\ref{fig:DE_mass_charge}, the dark electron's charge-to-mass ratio in the region of the parameter space we are interested is given by values $e_\chi/m_\chi \lesssim 10^{-14} \;e/m$. This would facilitate the formation of overdense regions with non-zero net charge, therefore allowing for the formation of dark-charged PBHs. The details and probabilities of this formation mechanism requires further study and we will leave this to future work.

\paragraph{Observational constraints} 
While the goals and directions of the present article were guided by observations, more work is needed to comprehensively evaluate the scenario we have considered here within the full breadth of efforts that were and are currently undertaken to constrain dark forces, PBHs and dark matter. In the following, we attempt to take a first step toward linking our considerations more closely to these pluriform undertakings.

Given the low mass range $(M_\text{PBH}\lesssim 10^{-15}M_\odot)$ for the primordial black holes proposed in this article, usual tests focusing on lensing and dynamical effects (dynamical friction, disk heating and PBH interaction with stars and local objects) will not be useful in placing tighter constraints on $f_\textrm{PBH}$ (see \cite{Carr:2020gox}). Gravitational waves, on the other hand, are still a viable route to do so. According to the latest LISA review of PBHs' gravitational wave signatures~\cite{LISACosmologyWorkingGroup:2023njw}, the estimated sensitivity for LISA will be able to probe this mass range, although this will not be located in their peak sensitivity region as shown in their figure 23; other upcoming space-based detectors, such as TianQin and Taiji, have slightly increased sensitivity in the corresponding frequency range \cite{Luo:2025ewp,Hu:2017mde}. Furthermore, given that the formation mechanisms (and post-reheating density) of these dark-charged PBHs might sufficiently vary from the standard cases, the stochastic background models based on the expected merging rates of PBHs would have to be updated. Moreover, the very presence of charge and a dark matter envelope around the black hole can also impact the gravitational wave signals being emitted (see for instance \cite{Croon:2017zcu,Alexander:2018qzg,Kopp:2018jom,Christiansen:2020pnv,Dror:2021wrl,Poddar:2023pfj,Bhalla:2024lta, Zhang:2024hrq,Barbosa:2025uau}). Regarding gravitational wave signals of charged black holes, an interesting analysis which can be directly extended to the model here presented can be found in \cite{Cardoso:2016olt}, where the authors have allowed the black holes to have a charge sourced by minicharged dark matter.

Taking a different approach, other routes to probe the dark-charged PBH model come from the emission of dark particles in various contexts. 
Some examples are: analyzing the impact on the thermal state and dynamics of the interstellar medium \cite{Takeshita:2024ytb, Wadekar:2022ymq},
looking for luminosity variations in accretion disks due to a dark-matter envelope around a black hole \cite{Boshkayev:2020kle}, considering the velocity distribution of dark matter spikes \cite{Zhang:2024hrq} and variations in rotation curves coming from introducing new forms of dark matter \cite{Faber:2005xc}. Finally, we note that extensions of this current work allowing for coupling between standard model and dark particles will open even broader probing mechanisms, including direct detection from experiments such as CDEX-10 \cite{CDEX:2024xqm}.

In short, there are many interesting observational channels that one might consider for developing phenomenological constraints.

\bigskip
\hrule\hrule\hrule

\vfill

\section*{Acknowledgments}
JS and JF acknowledge support from the Taiwan National Science and Technology Council, grant No. 112-2811-M-002-132. JF also acknowledges support from the European Union and Czech Ministry of Education, Youth and Sports through the FORTE project No. CZ.02.01.01/00/22\_008/0004632. SS is supported by a stipend for foreign postdoctoral researchers in Sweden of the Wenner Gren Foundations. 
MV was supported, during early phases of this project, by the Marsden Fund, via a grant administered by the Royal Society of New Zealand.

The authors wish to thank Lucas Cornetta and Anne Reinarz for their help with the numerics during the early stages of the project. 
The authors also wish to thank Emma Greenbank, Ilia Musco, Martin Spinrath, Jan Tristram Acuña, Harry Goodhew and Gordon Lee
for helpful discussions. 
JF thanks Will Barker for suggesting several references.
Finally, the authors thank the authors of \cite{Carr:2020gox} 
for permission to reuse figure~\ref{fig:PBH4}.

\appendix
\section{The specific heat of RN black holes}\label{app:spec-heat}
    
\begin{figure}[!ht]
        \centering
        \includegraphics[width=0.65\textwidth]{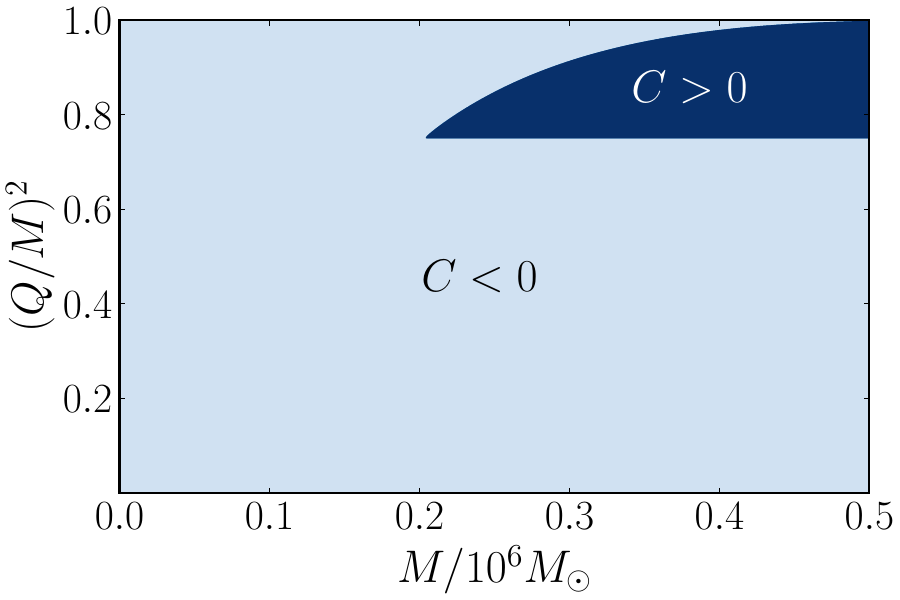}
        \caption{Recreation of figure~1 from \cite{Hiscock:1990ex} showing the positive and negative specific heat regions. As discussed in HW and in \cite{Ong:2019vnv}, the region of positive specific heat is always below the attractor curve.}
        \label{fig:HW1}
    \end{figure}

    Here we repeat another interesting result regarding the presence of a region (in configuration space) of positive specific heat for Reissner--Nordström black holes \cite{Davies:1977bgr,Hiscock:1990ex,Ong:2014dzm}. By definition, the specific heat of the given thermodynamical system is given by:
    \begin{equation}
	C = \frac{\dif M}{\dif T} = \frac{\dif M}{\dif t}\frac{\dif t}{\dif T}.\label{eq:C}
	\end{equation}
    For the case of a RN black hole, using~\eqref{eq:dMdtHWfull} and~\eqref{eq:Temp} we have:\footnote{As pointed out in \cite{Ong:2014dzm}, there is a small typo in equation~(23) of \cite{Hiscock:1990ex}. In our equation~\eqref{eq:dTdt} we present this equation with the typo corrected.}
    \begin{align}
		\frac{\dif T}{\dif t}&=\frac{e^{4}}{4 \pi^{4} m^{2}} \frac{Q^{4}}{r_{+}^{6}} \exp \left[\frac{-r_{+}^{2}}{Q Q_{0}}\right]\nonumber\\
        &\hphantom{=}-\frac{\hbar^{2} \alpha}{3840 \pi^2} \frac{\rbr{M^{2}-Q^{2}}^{3/2}\left(3 M+\sqrt{9 M^{2}-8 Q^{2}}\right)^{4}\left(M-2\sqrt{M^{2}-Q^{2}}\right)}{r_{+}^{10}\left(3 M^{2}-2 Q^{2}+M\sqrt{9 M^{2}-8 Q^{2}}\right)}.\label{eq:dTdt}
	\end{align}
       Given that $\dif M/ \dif t$ is always negative, one must simply set the LHS of~\eqref{eq:dTdt} equal to zero in order to be able to reproduce the specific heat behaviour. This sign change can then be derived with simple analytical estimates, which we will now do. The resulting behaviour of the specific heat $C$ can be seen in figure~\ref{fig:HW1}, a recreation of HW figure~1.

       The first term in equation~\eqref{eq:dTdt} is always positive. The second one is positive if $Q^2/M^2<3/4$, and negative if $Q^2/M^2>3/4$. As $\dif M/\dif t$ is always negative, $C$ will be positive only if $\dif T /\dif t$ is negative, which can thus only happen if $3/4 < Q^2/M^2 < 1$. The key point here is that $\left[M-2\left(M^{2}-Q^{2}\right)^{1 / 2}\right]$ can change sign.
       In order to check the region of parameter space where this happens, let us re-arrange and exchange $Q$ for $Y=Q^2/M^2$. This leads to the following condition for $C>0$:
 	\begin{equation}
 		\frac{960 e^2}{\pi^2 m^2 \hbar^2 \alpha} < \frac{M^2\rbr{1-Y}^{3/2}(3+\sqrt{9-8Y})^4(1-2\sqrt{1-Y})}{Y^2 r_+^4 (3 - 2 Y + \sqrt{9-8Y})}e^{r_+^2/(M\sqrt{Y}Q_0)}.\label{eq:10}
 	\end{equation}
 	The RHS's denominator can be bounded from above as:
 	\begin{equation}
 		Y^2 r_+^4 (3 - 2 Y + \sqrt{9-8Y}) < \rbr{\frac{9}{8}}^2\rbr{\sbr{1+\sqrt{12}}^2+1}M^{4}.
 	\end{equation}
 	Inserting this result into equation~\eqref{eq:10}, we have bound the full expression as
 	\begin{align}
         &\frac{M^8 \rbr{1-Y}^{3/2}(3+\sqrt{9-8Y})^4(1-2\sqrt{1-Y})}{M^{10} \frac{81}{64}([1+\sqrt{12}]^2+1)}\exp\rbr{\frac{\sbr{M\cbr{1+\sqrt{1-Y}}}^2}{M\sqrt{Y}Q_0}}\\
 		&\hspace{2cm} > \frac{16384}{81(14+4\sqrt{3})M^2}\rbr{1-Y}^{3/2}\rbr{1-2\sqrt{1-Y}}\exp\rbr{\frac{M\sbr{2+2\sqrt{1-Y}-Y}}{\sqrt{Y}Q_0}}\nonumber\\
 		&\hspace{2cm} > \frac{960 e^2}{\pi^2 m^2 \hbar^2 \alpha}.
 	\end{align}
 	Or, in full:
 	\begin{equation}
 		\frac{1215(7+2\sqrt{3})}{128}\frac{e^4}{\pi^2m^2\hbar^2 \alpha} < \frac{\rbr{1-Y}^{3/2}(1-2\sqrt{1-Y})}{M^2}\exp\rbr{\frac{M\sbr{2+2\sqrt{1-Y}-Y}}{\sqrt{Y}Q_0}}.
 	\end{equation}
    Given the dependence on $M$ on the RHS, it is fairly obvious now that for sufficiently large $Y$ and sufficiently large $M$ \emph{some} region of $C>0$ has to exist. Since, however, quite a few approximations were made, the numerical result of figure~\ref{fig:HW1} is more informative.

\paragraph{\texorpdfstring{Further insights into $\dif T/\dif t$}{Further insights into dT/dt}}
It is also help- and insightful to shorten equation~\eqref{eq:dTdt} somewhat, though this will limit (sometimes) the range of validity. For $Q^2 \leq(8/9) M^2$ there is an outer circular photon orbit at
\begin{equation}
    {r_\gamma}^+ = \oo{2} \left(3 M+\sqrt{9 M^{2}-8 Q^{2}}\right)\label{eq:rgammaplus}
\end{equation}
so
\begin{equation}
		\frac{\dif T}{\dif t}=\frac{e^{4}}{4 \pi^{4} m^{2}} \frac{Q^{4}}{r_{+}^{6}} \exp \left[\frac{-r_{+}^{2}}{Q Q_{0}}\right]
        -\frac{\hbar^{2} \alpha}{240 \pi^2} \frac{\sqrt{M^{2}-Q^{2}}^{\;3}
        \left({r_\gamma}^+\right)^{4}\left(3M-2r_+\right)}
        {r_{+}^{10}\left(3 M^{2}-2 Q^{2}+M\sqrt{9 M^{2}-8 Q^{2}}\right)},\label{eq:dTdt2}
\end{equation}
or even
\begin{align}
    \frac{\dif T}{\dif t}   &=\frac{e^{4}}{4 \pi^{4} m^{2}} \frac{Q^{4}}{r_{+}^{6}} \exp \left[\frac{-r_{+}^{2}}{Q Q_{0}}\right]
            -\frac{\hbar^{2} \alpha}{480 \pi^2} \frac{(M^{2}-Q^{2})^{\;3/2}\;
            \left({r_\gamma}^+\right)^{4}\left(3M-2r_+\right)}
            {r_{+}^{10}\left(M {r_\gamma}^+ - Q^{2}\right)}.\label{eq:dTdt3}
\end{align}
Both equations~\eqref{eq:dTdt2} and~\eqref{eq:dTdt3} can only be defined if ${r_\gamma}^+$ itself is well-defined.

More drastically, the impact parameter corresponding to the outer circular photon orbit is given by \cite{Capobianco:2019thesis}
\begin{equation}
b_+^2 = \frac{({r_\gamma}^+)^4}{ \frac{3}{2} M^{2} (1-\frac{2Q^2}{3 M^2} +\sqrt{1-\frac{8Q^2}{9M^2}})}
     = \frac{ ({r_\gamma}^+)^4}{M{r_\gamma}^+ - Q^2}\;,
\end{equation}
allowing us to write
\begin{align}
\frac{\dif T}{\dif t}&=\frac{e^{4}}{4 \pi^{4} m^{2}} \frac{Q^{4}}{r_{+}^{6}} \exp \left[\frac{-r_{+}^{2}}{Q Q_{0}}\right]
        -\frac{\hbar^{2} \alpha}{480 \pi^2} \frac{(M^{2}-Q^{2})^{\;3/2}\;
        b_+^{2}\left(3M-2r_+\right)}
        {r_{+}^{10}}\;.
\end{align}

    \section{Numerics}
    \label{S:numerics}

     \subsection{Numerical implementation}
    We have implemented equations~\eqref{eq:dmudtHWnew} and~\eqref{eq:dYdtHWnew} in the Julia language. Our implementation makes use of \texttt{v6.82.0} of the library \texttt{OrdinaryDiffEq.jl}, a component package of \texttt{DifferentialEquations.jl} \cite{SciMLDiffEq:2017}; \texttt{OrdinaryDiffEq.jl} library contains a state-of-the-art integration suite featuring integration methods with adaptive timestepping and stiffness detection, as well as auto-switching algorithms that automatically switch from an efficient solver to a stiff solver when the system in question becomes stiff. \texttt{OrdinaryDiffEq.jl} includes an integrated testing suite (including convergence testing) for validation of the integration methods. Following the recommendations in the documentation \cite{SciMLDiffEq:2017} for \texttt{DifferentialEquations.jl}, our implementation makes use of the solver \texttt{autoTsit5(Rosenbrock23)}, which employs an order 5/4 Tsitouras Runge-Kutta method \cite{Tsitouras:2011} as the primary integration method, and an order 2/3 Rosenbrock-W method \cite{Rosenbrock:1963,Press:2007numerical} when the system becomes stiff; we verified that for \texttt{v6.82.0}, the aforementioned methods passed unit tests (including convergence tests) found in the supplied testing suite. The numerical codes and scripts can be found on the Github repository \url{https://github.com/justincfeng/bhevapsolver}.

\subsection{How far are we from extremality?} \label{app:extremality}
\begin{figure}[!ht]
    \centering
    \includegraphics[width=0.65\linewidth]{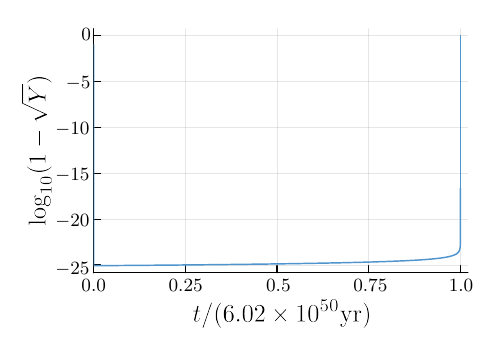}
	\caption{Plot of $1-\sqrt{Y}$ for the numerical solution that most closesly approaches extremality, corresponding to  $\xi=10^{-14}$ and $\mu_\mathrm{h}=0.90$. The near-extremal mass for this solution is on the order of $\sim10^{-17}M_{\odot}$.} 
	\label{fig:dYMin}
    \end{figure}

 {The numerical solutions allow us to verify that the condition $M-Q \gg m_{\textrm{Planck}}^4/Q^3$, (cf. the inequality at the beginning of Sec. 3 in \cite{Page:2000dk}), holds throughout the evolution. This condition, which can be rewritten as $Y^{3/2}(1-\sqrt{Y}) \gg (m_{\textrm{Planck}}/M)^4$, is needed to avoid the onset of effects arising from the discreteness of states near extremality \cite{Preskill:1991tb,Maldacena:1998uz,Page:2000dk,Brown:2024ajk}. In our numerical solutions, we find that the charge-to-mass ratio $\sqrt{Y}$ never exceeds $\sqrt{Y}\sim 1-10^{-25}$. We illustrate this for the most extreme case we encountered in Fig.~\ref{fig:dYMin}. The smallest masses considered are $10^{-25} M_{\odot}\sim 10^{13}m_{\textrm{Planck}}$, so that the maximum threshold is $(m_{\textrm{Planck}}/M)^4\sim 10^{-52}$.} It follows that the solutions we have considered here remain well above the threshold. 

\section{Geometrodynamic units}\label{app:units}

{Both in the current article as well in in HW \cite{HiscockWeems1990} geometrodynamic units were adopted. This means that everything is ultimately expressed in units of length and $G=c=1$. A table containing conversion factors between SI and geometrodynamic units can be found in the Appendix section of \cite{Wald:1984rg}. 
Below, we list the main constants and parameters used by us and HW and their respective geometrodynamic units dimensions for clarity:
\begin{equation}
    [q_e] = [L]\;, [m_e] = [L]\;, [\hbar] = [L^2]\;, [Q_0] = [L]\;.
\end{equation}
The conversion of mass to meters is given by the factor $G/c^2$. The elementary charge $q_e$ is given in units of Coulombs and one can convert it to meters by multiplying the charge by a factor of $\sqrt{G}/(c^2\sqrt{4\pi\varepsilon_0})$. For completeness, to convert $\hbar$ from {\sisetup{inter-unit-product = \ensuremath { { } \cdot { } } }\si{\joule\second}} to \si{\metre\squared}, we multiply it by a factor of $G/c^3$. Below, we present the values of the main quantities used in this article:
    \begin{align}
        M_\odot &= \SI{1477}{\metre}\;,\nonumber\\
	Q_0 &= \SI{2.5d8}{\metre}\;,\nonumber\\
	\hbar &= \SI{2.6d-70}{\metre\squared}\;,\\
	m_e &= \SI{6.75d-58}{\metre}\;,\nonumber\\
	e &= \SI{1.38d-36}{\metre}\;.\nonumber
    \end{align}}

\hrule\hrule\hrule

\addcontentsline{toc}{section}{References}
\printbibliography

\end{document}